\documentclass[twocolumn,showpacs,superscriptaddress,prl,aps,floatfix]{revtex4-1}
\usepackage{graphicx}
\usepackage{rotating}
\usepackage{amsmath}
\usepackage[usenames,dvipsnames]{color}
\usepackage{float}
\usepackage{rotating}
\usepackage{booktabs}
\usepackage{multirow}
\usepackage{comment}
\usepackage[colorlinks,linkcolor=blue,citecolor=blue]{hyperref}
\usepackage[T1]{fontenc}
\usepackage{bm}
\usepackage{setspace}
\usepackage{xr-hyper}
\usepackage[colorlinks,linkcolor=blue,citecolor=blue]{hyperref}
\usepackage{xr}  
\makeatletter

\newcommand{\Rmnum}[1]{\expandafter\@slowromancap\romannumeral #1@}

\hfuzz=\maxdimen
\begin{document}
\title{Hidden Chiral Ferroelectricity in AgNbO$_3$ Perovskite}

\author{Ying Song}
\affiliation{Key Laboratory of Advanced Materials and Devices for Post-Moore Chips, Ministry of Education, University of Science and Technology Beijing, Beijing 100083, China}
\affiliation{Beijing Key Laboratory for Magneto-Photoelectrical Composite and Interface Science, School of Mathematics and Physics, University of Science and Technology Beijing, Beijing 100083, China}

\author{Lingzhi Cao}
\affiliation{Key Laboratory of Advanced Materials and Devices for Post-Moore Chips, Ministry of Education, University of Science and Technology Beijing, Beijing 100083, China}
\affiliation{Beijing Key Laboratory for Magneto-Photoelectrical Composite and Interface Science, School of Mathematics and Physics, University of Science and Technology Beijing, Beijing 100083, China}

\author{Jinming Zhai}
\affiliation{Key Laboratory of Advanced Materials and Devices for Post-Moore Chips, Ministry of Education, University of Science and Technology Beijing, Beijing 100083, China}
\affiliation{Beijing Key Laboratory for Magneto-Photoelectrical Composite and Interface Science, School of Mathematics and Physics, University of Science and Technology Beijing, Beijing 100083, China}

\author{Zhilong Yang}
\affiliation{Key Laboratory of Advanced Materials and Devices for Post-Moore Chips, Ministry of Education, University of Science and Technology Beijing, Beijing 100083, China}
\affiliation{Beijing Key Laboratory for Magneto-Photoelectrical Composite and Interface Science, School of Mathematics and Physics, University of Science and Technology Beijing, Beijing 100083, China}

\author{Yali Yang}
\email{ylyang@ustb.edu.cn}
\affiliation{Key Laboratory of Advanced Materials and Devices for Post-Moore Chips, Ministry of Education, University of Science and Technology Beijing, Beijing 100083, China}
\affiliation{Beijing Key Laboratory for Magneto-Photoelectrical Composite and Interface Science, School of Mathematics and Physics, University of Science and Technology Beijing, Beijing 100083, China}

\author{Laurent Bellaiche}
\affiliation{Smart Ferroic Materials Center, Physics Department and Institute for Nanoscience and Engineering, University of Arkansas, Fayetteville, Arkansas 72701, USA}
\affiliation{Department of Materials Science and Engineering, Tel Aviv University, Ramat Aviv, Tel Aviv 6997801, Israel}

\author{Jiangang He}
\email{jghe2021@ustb.edu.cn}
\affiliation{Key Laboratory of Advanced Materials and Devices for Post-Moore Chips, Ministry of Education, University of Science and Technology Beijing, Beijing 100083, China}
\affiliation{Beijing Key Laboratory for Magneto-Photoelectrical Composite and Interface Science, School of Mathematics and Physics, University of Science and Technology Beijing, Beijing 100083, China}


\date{\today}

\begin{abstract}
AgNbO$_3$ is a  lead-free perovskite with considerable potential for energy storage and optoelectronic applications, yet its low-temperature crystal structure has remained controversial. In this Letter, we revisit its low-energy structural landscape using a systematic first-principles structural search based on symmetry-adapted phonon-mode theory. We uncover a previously unreported chiral ferroelectric phase with space group $R3$, which exhibits a large spontaneous polarization and a low polarization switching barrier, enabling polarization reversal under electric fields. Crucially, the structural chirality of this phase is intrinsically locked to the ferroelectric polarization, allowing electrical control of the chiral handedness. Consequently, chiral optical responses—including circular dichroism, circular photogalvanic effect, optical activity, and second-order nonlinear optics—can be reversibly switched by an external electric field. These results not only clarify the complex low-temperature structural behavior of AgNbO$_3$ but also establish a rare purely inorganic platform for electric-field-tunable chirality, opening a pathway toward ultrafast, electrically controlled chiral optoelectronics.
\end{abstract}


\maketitle


AgNbO$_3$ is a rare example of lead-free antiferroelectric perovskites~\cite{guo2025synergistic,yang2020lead,ting2023advances,xue2025superior,zhang2021agnbo3,ma2022dielectric}, and has attracted sustained interest owing to its potential applications in energy storage and conversion, as well as in photocatalysis and piezoelectric devices~\cite{tian2022silver,D0TA08345C}. Despite extensive experimental and theoretical efforts, the crystal structure of AgNbO$_3$ remains highly controversial, particularly at low temperatures~\cite{sciau2004structural}. Reference~\cite{Ph_Sciau_2004} indicated that, upon cooling from 852~K to 1.5~K, AgNbO$_3$ undergoes a sequence of structural phase transitions: from a cubic phase ($Pm\bar{3}m$) to a tetragonal phase ($P4/mbm$), followed by an orthorhombic phase ($Cmcm$), and subsequently three distinct orthorhombic phases commonly denoted as $M_1$, $M_2$, and $M_3$. Among them, the $M_1$ phase has been identified as ferroelectric with space group $Pmc2_1$~\cite{doi:10.1021/cm103389q}, whereas the $M_2$ and $M_3$ phases are antiferroelectric and share the same space group $Pbcm$. The proposed $Pmc2_1$ structure can account for the experimentally observed antiferroelectric-like hysteresis loop with a large polarization of 52~$\mu$C/cm$^2$ in polycrystalline samples subjected to high electric fields up to 220~kV/cm~\cite{10.1063/1.2751136}. However, subsequent first-principles calculations questioned the existence of the $Pmc2_1$ phase, as no unstable phonon mode of $Pbcm$ was found to connect to this structure~\cite{Moriwake_2012}. Later experiments by Tian \emph{et al.} reported ferroelectric-like domain switching at low electric fields up to 340~K, together with the observation of submicron polar regions by transmission electron microscopy~\cite{C6TA06353E}, lending renewed support to the presence of a polar phase. In addition, first-principles studies predicted a rhombohedral $R3c$ phase as the ground-state structure of AgNbO$_3$ at 0~K~\cite{AgNbO3-R3c}, as it is nearly degenerate in energy with the experimentally observed $Pbcm$ phase~\cite{zhang2024lattice}.

Beyond switchable polarization, ferroelectrics (or antiferroelectricity) exhibit strong coupling between structural distortions and other order parameters, including lattice strain~\cite{PhysRevLett.17.198,kundys2010light,10.1063/1.4905505}, electronic structure~\cite{doi:10.1021/jacs.4c03296}, optical responses~\cite{doi:10.1021/jacs.4c03296,doi:10.1021/jacs.4c13604}, and magnetism in some cases~\cite{fiebig2016evolution,huang2024manipulating}. Such couplings enable electric-field control over a broad range of physical properties. For instance, coupling between polarization and spin degrees of freedom can give rise to electrically switchable Rashba spin splitting~\cite{https://doi.org/10.1002/adma.201203199,doi:10.1021/acs.nanolett.7b04829,liebmann2016giant,he2018tunable} and strong magnetoelectric effects in antiferroelectric altermagnets~\cite{PhysRevLett.134.106801}.

In parallel, chiral materials~\cite{niu2023chiral,Bousquet_2025} have attracted increasing attention due to their unique roles in nonlinear optics (NLO)~\cite{powers2013field}, quantum optics~\cite{nussenzveig1973introduction}, magneto-optical effects~\cite{doi:10.1073/pnas.56.5.1391}, and spintronics~\cite{vzutic2004spintronics}. The ability to interchange right- and left-handed chiral structures using external fields is crucial for controlling chirality-related functionalities. Chiral ferroelectrics, in particular, offer the appealing possibility of electrically tuning chiral-dependent phenomena such as NLO and circular photogalvanic effects (CPGE)~\cite{le2021topology,de2017quantized,le2020ab}. Despite this promise, the coexistence of chirality and ferroelectricity (or antiferroelectricity) is exceedingly rare. To date, experimentally realized chiral ferroelectrics are largely limited to hybrid organic--inorganic perovskites, in which chiral organic molecules transfer their handedness to the inorganic framework~\cite{https://doi.org/10.1002/smll.201902237,doi:10.1021/jacs.4c13604,DANG2021794}. By contrast, intrinsic chiral ferroelectricity in purely inorganic compounds remains exceptionally uncommon~\cite{luo2025strainin,gutierrezamigo2025}. An early study reported a hysteresis loop of natural optical activity (NOA) as a function of electric field in the ferroelectric compound Pb$_5$Ge$_3$O$_{11}$~\cite{10.1063/1.1653848}, whose chirality was recently shown to originate from a polar unstable mode of the paraelectric phase~\cite{PhysRevB.109.024113}.

In this Letter, we revisit the structural landscape of AgNbO$_3$ through a systematic structure search based on symmetry-adapted modes associated with the unstable phonons of the cubic perovskite phase. We identify a previously unreported phase with space group $R3$ that is energetically degenerate with the putative ground state. This $R3$ structure exhibits a large spontaneous polarization of approximately 50~$\mu$C/cm$^2$, nearly twice that of the archetypal ferroelectric perovskite BaTiO$_3$~\cite{PhysRev.99.1161}, together with a remarkably low polarization switching barrier of about 5~meV/atom. Crucially, the structural chirality in this phase is intrinsically locked to the spontaneous polarization, enabling electrical switching of the chiral handedness. As a consequence, chiral observables such as circular dichroism (CD), CPGE, NOA, and NLO can be controlled by an external electric field. Our results not only shed new light on the long-standing structural controversies of AgNbO$_3$ but also establish a viable pathway toward intrinsically inorganic, electrically switchable chiral photoelectronic materials.

Most density functional theory (DFT) calculations in this work, including structural relaxations and phonon analyses, were carried out using the Vienna \textit{ab initio} simulation package (VASP)~\cite{vasp1,vasp2}. The projector augmented-wave (PAW) method~\cite{paw1,paw2} was employed in combination with a plane-wave basis set with a kinetic energy cutoff of 520~eV. Exchange--correlation effects were treated using the Perdew--Burke--Ernzerhof functional revised for solids (PBEsol)~\cite{PhysRevLett.100.136406}. Further computational details are provided in the Supplemental Information (SI)~\cite{si}.

\begin{figure}[htbp]
	\centerline{
	\hspace*{0.1cm}\includegraphics[width=1\linewidth]{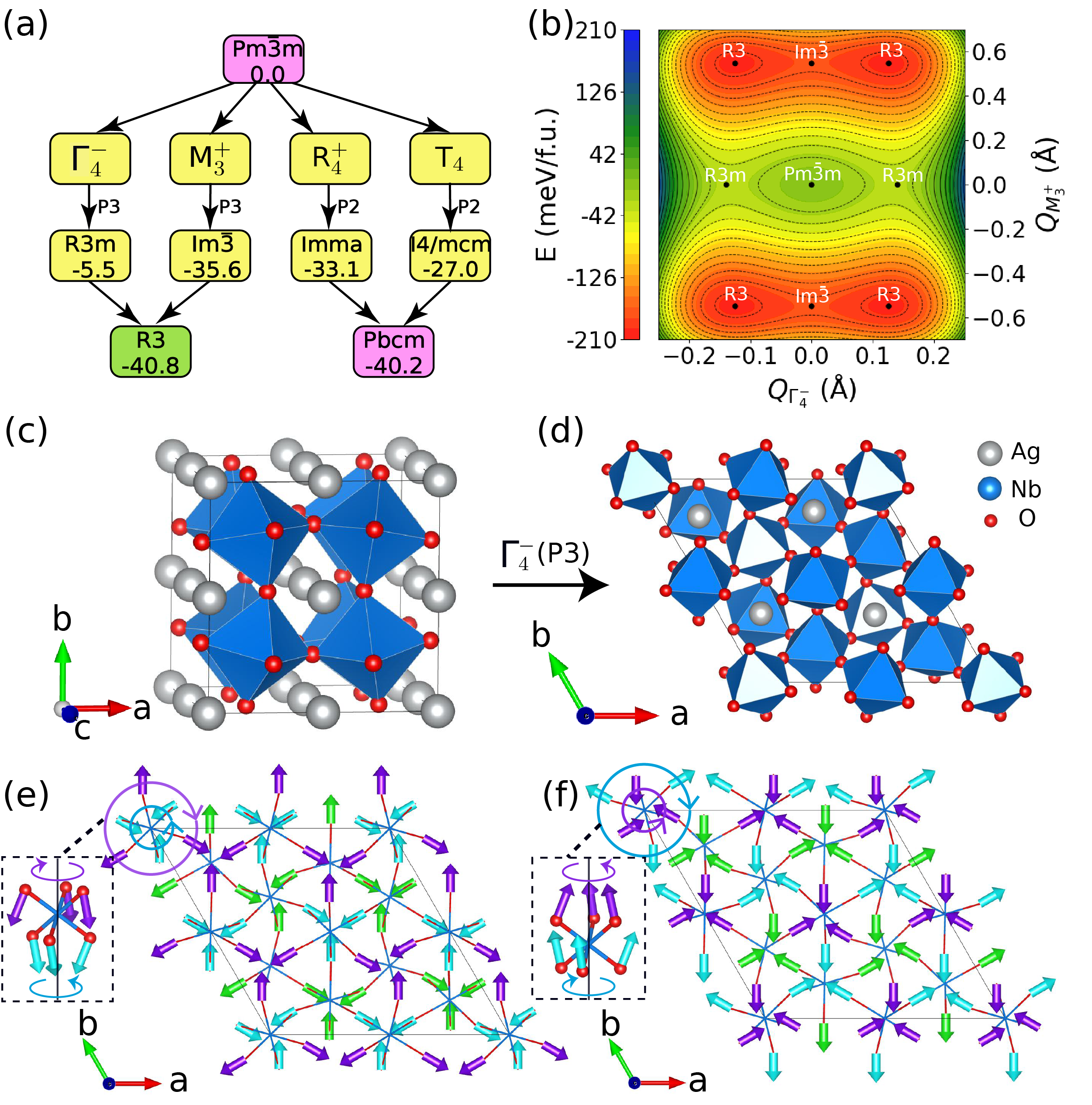}}
	\caption{(a) Group--subgroup tree highlighting the relationship of $Pbcm$ and $R3$ with respect to $Pm\bar{3}m$. (b) Free-energy contour map of the coupling between the $\Gamma_{4}^{-}$ and $M_{3}^{+}$ modes, see Table~\textcolor{magenta}{S2} for the fitting parameters. (c,d) Crystal structures of $Im\bar{3}$ and $R3$, respectively. (e,f) Atomic displacement patterns of the $\Gamma_{4}^-$ mode for positive and negative polarizations, respectively. Cyan, green, and purple arrows denote the in-plane ($ab$-plane) projections of the displacements of O atoms in different layers. Curved arrows indicate the sense of rotation of the O atoms. The insets show the vibrational components of the O atoms along the $c$ axis. Clockwise and counterclockwise rotations of the O atoms about the $c$ axis correspond to left- and right-handedness, respectively.}
	\label{fig:fig1}
\end{figure}

The phonon dispersion of the cubic AgNbO$_3$ at 0 K is shown in Fig.~\textcolor{magenta}{S1(a)}. The cubic phase has multiple unstable phonons with the large unstable phonon modes being observed at the $\Gamma$, M, R, and X points. The corresponding symmetry (with the language of irreducible representation, irrep) of the lowest phonon mode are $\Gamma_{4}^{-}$, M$_{3}^{+}$, R$_{4}^{+}$, and X$_5^-$, respectively. Other unstable phonon modes include T$_{4}$ (T is a line between M and R points) and $\Delta_{5}$ ($\Delta$ is a line between $\Gamma$ and X points). The second most unstable modes at $\Gamma$ and M points are $\Gamma_{5}^{-}$ and M$_{3}^{-}$, respectively. The atomic vibrations of these modes are illustrated in Fig.~\textcolor{magenta}{S2}. These unstable phonon modes mainly involve oxygen atoms. The subgroup structures were generated by displacing atom positions of the cubic phase based on the force constant eigenvectors of individual phonon modes, as well as combinations of two or three unstable phonon modes. Consequently, around 200 subgroup structures are generated. Each obtained subgroup structure then underwent full structural relaxation. The distortion amplitude of the fully relaxed subgroup structures relative to the cubic phase was computed using the \textsc{AMPLIMODES} software implemented in Bilbao Crystallographic Server~\cite{Perez-Mato:sh5107}, and the direction of the order parameter (identified via the space group symmetries of the parent and subgroup structures in the phase transition) was obtained using the \textsc{ISOTROPY} software~\cite{isotropy}. The main result of the group-subgroup tree is shown in Fig.~\ref{fig:fig1} and Fig.~\textcolor{magenta}{S1}(b), while the complete results of the subgroup structures, energies, and distortion amplitudes are tabulated in Table~\textcolor{magenta}{S1}.

As shown in Fig.~\ref{fig:fig1}, the antiferroelectric $Pbcm$ (No.~57) phase can be generated by the simultaneous condensation of the R$_4^+$ and T$_4$ modes along the P2(1,1,0) and P2(1,1,0,0,0,0) directions, respectively. The energy gain of the $Pbcm$ phase relative to the cubic $Pm\bar{3}m$ structure is 40.2~meV/atom, in excellent agreement with previous first-principles results~\cite{zhang2024lattice}. The fully relaxed lattice parameters are $a=5.5143$~\AA, $b=5.5976$~\AA, and $c=15.5072$~\AA, which closely match the experimental values ($a=5.5380$~\AA, $b=5.5930$~\AA, and $c=15.6200$~\AA)~\cite{sciau2004structural}. The low-energy rhombohedral $R3c$ structure, formed by the coupling of the $\Gamma_4^-$ mode along P3(1,1,1) and the R$_4^+$ mode along P3(1,1,1), lies 40.1~meV/atom below the $Pm\bar{3}m$ phase. This phase has been predicted in earlier first-principles studies~\cite{AgNbO3-R3c,zhang2024lattice}, although it has not yet been observed experimentally. Coupling the $\Gamma_4^-$ mode along P2(1,1,0), the M$_3^+$ mode along P1(1,0,0), and the R$_4^+$ mode along P2(0,1,1) yields an orthorhombic $Pmc2_1$ phase [see Fig.~\textcolor{magenta}{S1}(b)], which shares the same space group as the previously proposed $Pmc2_1$ phase~\cite{Moriwake_2012,tian2022silver}. Our group-theoretical analysis demonstrates that this $Pmc2_1$ phase is not a subgroup of $Pbcm$. Specifically, the $Pbcm$ structure is primarily constructed from the $\Delta_5$, R$_4^+$, and T$_4$ modes, whereas the distortion amplitudes of the $\Delta_5$ and T$_4$ modes vanish in the $Pmc2_1$ phase. Furthermore, the existence of this $Pmc2_1$ phase, together with the absence of any unstable phonon mode connecting the $Pmc2_1$ and $Pbcm$ phases, provides additional support for the conclusion that $Pmc2_1$ is not a subgroup of $Pbcm$. The spontaneous polarization of this $Pmc2_1$ phase, calculated using the centrosymmetric $Imma$ phase as the reference, is approximately 54~$\mu$C/cm$^{2}$, in good agreement with the experimentally reported value~\cite{10.1063/1.2751136}.

Interestingly, we find that the lowest-energy structure calculated using PBEsol~\cite{PhysRevLett.100.136406} is in fact the rhombohedral $R3$ (No.~146) phase [see Fig.~\ref{fig:fig1}(b)], which is 40.8~meV/atom lower in energy than the $Pm\bar{3}m$ phase and that is stabilized by the coupling of the $\Gamma_4^-$ and M$_3^+$ modes along the P3(1,1,1) direction. Calculations with other exchange--correlation functionals confirm that the $R3$ phase either has the lowest energy or is nearly degenerate with the $Pbcm$ phase (see Table~\textcolor{magenta}{S3}). The potential energy surface expanded by $\Gamma_4^-$ and M$_3^+$ modes are shown in Fig.~\ref{fig:fig1}(b), which demonstrates a clear coupling between these two modes.
Phonon calculations further demonstrate that the $R3$ phase is dynamically stable at 0~K, as evidenced by the absence of imaginary phonon frequencies (see Fig.~\textcolor{magenta}{S4}). We further assessed its thermodynamic stability by evaluating the Gibbs free energy, $G = U - TS + pV = H - TS$, at room temperature within the harmonic approximation, where $U$, $T$, $S$, $p$, and $V$ denote the total energy, absolute temperature, vibrational entropy, external pressure, and volume, respectively. At ambient pressure, the $pV$ contribution is negligible for solids; therefore, we approximate the Gibbs free energy by the Helmholtz free energy, $H = U - TS$. As shown in Fig.~\textcolor{magenta}{S5}, the PBEsol calculated Helmholtz free energy of the $R3$ phase remains lower than that of the $Pbcm$ phase at room temperature, although the energy difference is reduced. We therefore conclude that the thermodynamic stability of the $R3$ phase is at least comparable to that of the $Pbcm$ phase.

\begin{figure}[htbp]
	\centering
	\includegraphics[width=1\linewidth]{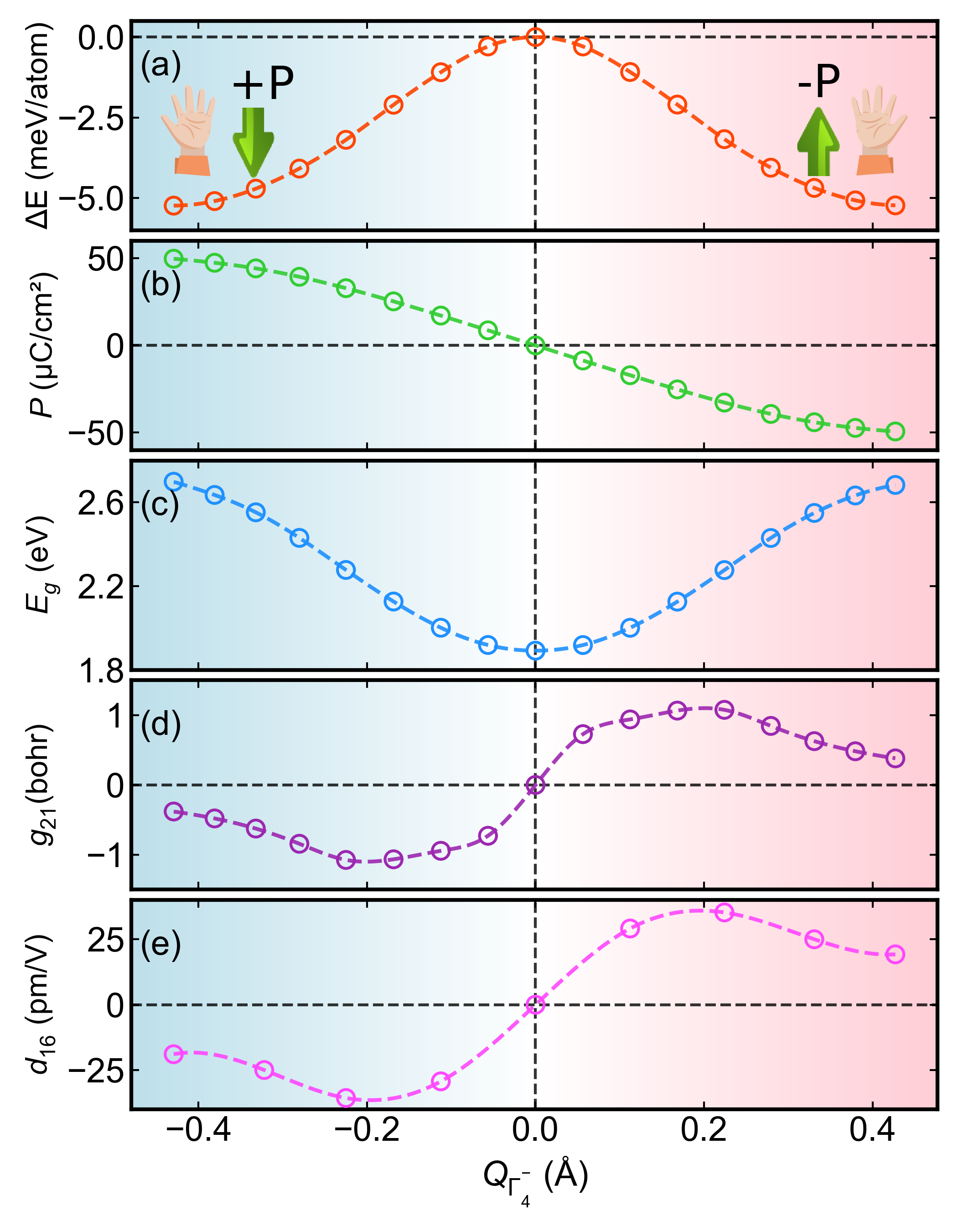}
	\caption{Evolution of (a) the energy difference $\Delta E$ between the $R3$ and $Im\bar{3}$ phases, (b) the spontaneous polarization $P$, (c) the band gap $E_{\mathrm{g}}$, (d) the largest gyration-tensor component $g_{21}$ associated with natural optical activity (NOA), and (e) the largest second-order nonlinear-optical coefficient $d_{16}$, as functions of the amplitude of the $\Gamma_{4}^{-}$ mode, $Q_{\Gamma_{4}^{-}}$, which connects the $R3$ and $Im\bar{3}$ phases.}
	\label{fig:fig2}
\end{figure}

Because the $R3$ phase [Fig.~\ref{fig:fig1}(d)] belongs to a polar space group, we examine its ferroelectric character. As shown in Fig.~\ref{fig:fig1}(a), the centrosymmetric $Im\bar{3}$ phase [see Fig.~\ref{fig:fig1}(c)], obtained by displacing atoms along the force-constant eigenvector of the M$_3^+$ mode in the P3(1,1,1) direction, exhibits a significantly larger energy lowering ($-35.6$~meV/atom) relative to the cubic $Pm\bar{3}m$ phase than the $R3m$ phase produced by condensing the $\Gamma_4^-$ mode along the same direction. Therefore, the lowest energy pathway to switch the direction of the polarization is along the $\Gamma_4^-$, see Fig.~\ref{fig:fig1}(b). This establishes $Im\bar{3}$ as the appropriate centrosymmetric reference structure for the polar $R3$ phase. We computed the phonon dispersion of the $Im\bar{3}$ phase and identified a strongly unstable polar mode at the $\Gamma$ point with $\Gamma_{4}^-$ symmetry (Fig.~\textcolor{magenta}{S6}). This one-dimensional order parameter directly connects the $Im\bar{3}$ and $R3$ phases, as illustrated in Fig.~\ref{fig:fig1}(a) and (b). Condensation of the $\Gamma_{4}^-$ mode along the P3(1,1,1) direction lowers the total energy by approximately 5~meV/atom, thereby stabilizing the polar $R3$ structure. The corresponding eigenvector, shown in Figs.~\ref{fig:fig1}(e) and (f), consists of oxygen rotations within the $ab$ plane accompanied by collective displacements along the $c$ axis, giving rise to chiral and polar symmetry at the same time. Importantly, the rotational and translational oxygen motions are symmetry-locked. The Nb atoms primarily displace along the $c$ axis (Fig.~\textcolor{magenta}{S7}).

\begin{figure}[htbp]
	\includegraphics[width=1\linewidth]{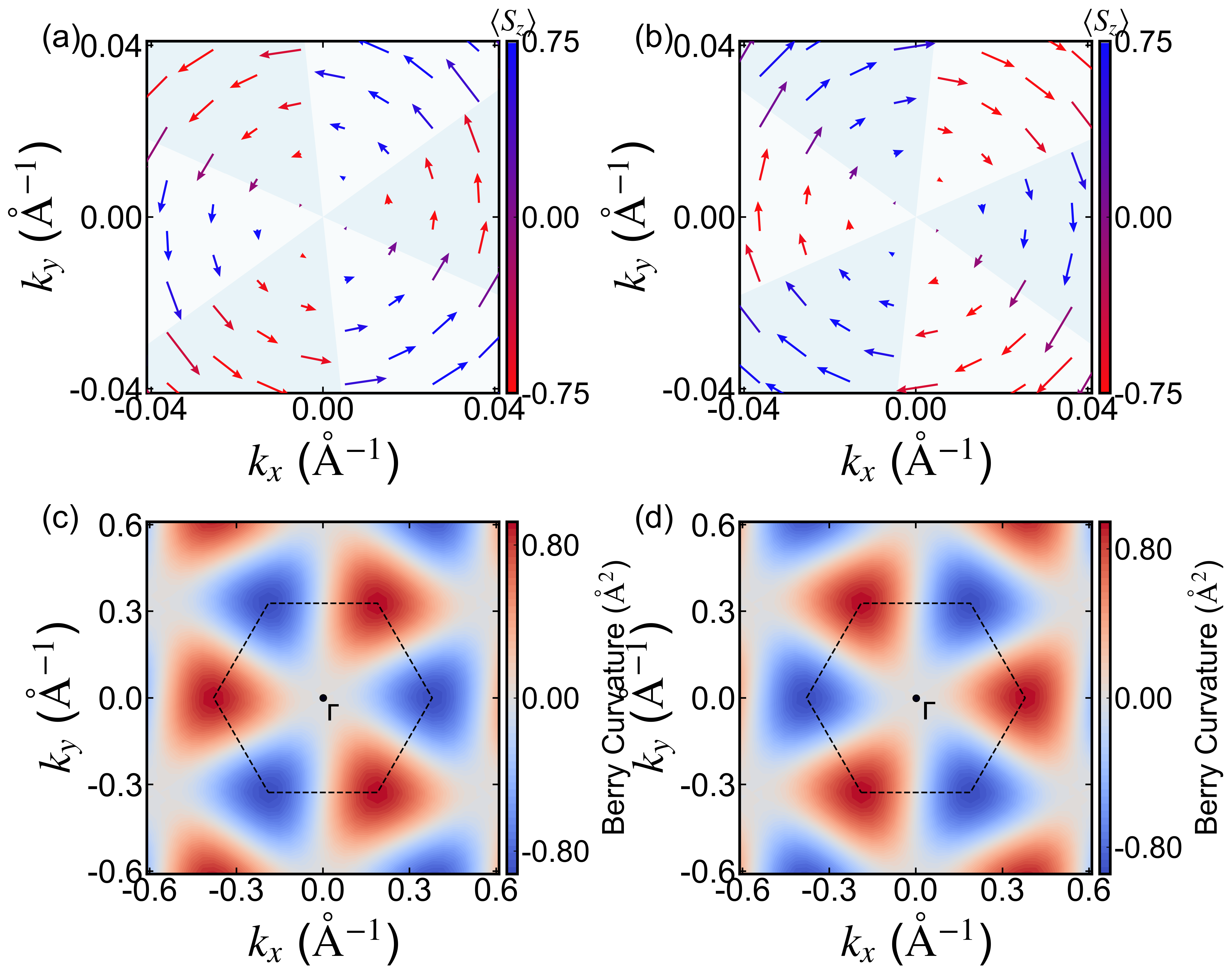}
	\caption{(a,c) Spin texture of the lowest conduction band near the $\Gamma$ point and the momentum-resolved Berry curvature for left-handed ($+$ polarization) $R3$ AgNbO$_3$, respectively. (b,d) Spin texture of the lowest conduction band near the $\Gamma$ point and the momentum-resolved Berry curvature for right-handed ($-$ polarization) $R3$ AgNbO$_3$, respectively.}
	\label{fig:fig3}
\end{figure}

The minimum-energy path on the potential energy surface connecting the two $R3$ structures with opposite polarizations via the centrosymmetric $Im\bar{3}$ phase is investigated using the climbing-image nudged elastic band (cNEB) method~\cite{10.1063/1.1329672}. As shown in Fig.~\ref{fig:fig2}(a), the calculated path exhibits a low energy barrier ($\Delta E$) of 5~meV/atom between the two oppositely polarized $R3$ phases. The transition state corresponds to the $Im\bar{3}$ phase, in full agreement with the phonon calculations. The total atomic displacement between the $R3$ and $Im\bar{3}$ phases is 0.43~\AA, which is comparable to that of archetypal displacive ferroelectrics (see Fig.~\textcolor{magenta}{S8}). Figure~\ref{fig:fig2}(b) clearly shows that the two polar structures possess spontaneous polarizations of equal magnitude and opposite sign (approximately $\pm 50$\,\textmu C/cm$^{2}$), whereas the spontaneous polarization of the $Im\bar{3}$ phase is strictly zero. Owing to the relatively small $\Delta E$ and the sizable polarization, the estimated electric field required for polarization switching is about 14~MV/cm, which is comparable to that of typical ferroelectric materials such as BaTiO$_3$ and PbTiO$_3$ (see Fig.~\textcolor{magenta}{S8}). These results indicate that the spontaneous polarization in $R3$ AgNbO$_3$ can be switched by an experimentally feasible electric field.

Interestingly, the electronic band gap (E$_{\mathrm{g}}$) of AgNbO$_3$ calculated using the HSE06 functional~\cite{10.1063/1.1564060} increases significantly from the $Im\bar{3}$ phase (1.88~eV) to the $R3$ phase (2.70~eV), as shown in Fig.~\ref{fig:fig2}(c). Such behavior is relatively rare among ferroelectric materials and has been reported only in a limited number of systems~\cite{he2018tunable,doi:10.1021/jacs.4c03296}, suggesting a strong coupling between the electronic structure and the polar distortion. This enhancement of E$_{\mathrm{g}}$ can be attributed to the simultaneous downward shift of the valence-band maximum and upward shift of the conduction-band minimum. The valence-band states are mainly derived from filled antibonding states formed by Ag-4$d$ and O-2$p$ hybridization, while the conduction-band states originate primarily from empty antibonding states associated with Nb-4$d$ and O-2$p$ hybridization. These changes are induced by the Jahn--Teller distortion driven by the $d^{0}$ electronic configuration of the Nb$^{5+}$ cation (see Fig.~\textcolor{magenta}{S9}). The correlation between the Nb--O bond length and the $\Gamma_{4}^{-}$ distortion is illustrated in Fig.~\textcolor{magenta}{S10}. Larger Nb--O bond lengths and O--Nb--O bond angles weaken the Nb-$d$--O-$p$ hybridization, leading to progressively narrower bands near the Fermi level and a concomitant increase in E$_{\mathrm{g}}$. The conduction-band minimum of $R3$ AgNbO$_3$ is located at the T point of the first Brillouin zone, which lies only 6 meV below the second-lowest conduction-band minimum at the $\Gamma$ point. Notably, a pronounced Rashba spin splitting is observed at the $\Gamma$ point (see Fig.~\textcolor{magenta}{S11}). The corresponding spin textures of the $R3$ structures with positive and negative polarizations are shown in Figs.~\ref{fig:fig3}(a) and (b), respectively, demonstrating a clear correlation between spin texture and polarization direction, consistent with observations in other ferroelectric materials~\cite{https://doi.org/10.1002/adma.201203199,he2018tunable}.

\begin{figure}[htbp]
	\centering
	\includegraphics[width=1\linewidth]{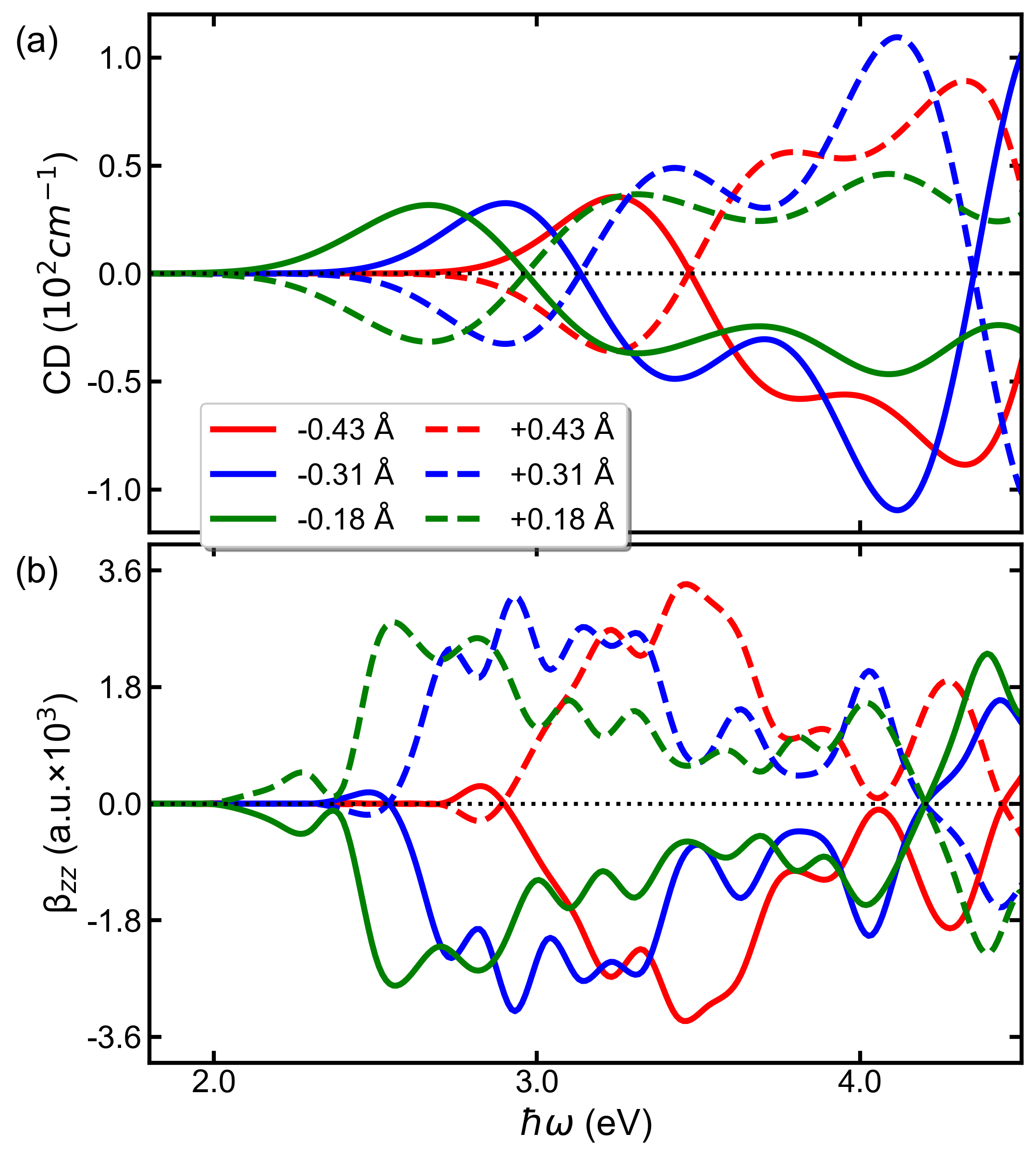}
	\caption{(a) Circular dichroism (CD) spectra of the $R3$ phase for different distortion amplitudes of the polar $\Gamma_{4}^{-}$ mode. (b) Circular photogalvanic effect (CPGE) tensor component ($\beta_{\mathrm{zz}}$) of the $R3$ phase as a function of the distortion amplitude of the polar $\Gamma_{4}^{-}$ mode.}
	\label{fig:fig4}
\end{figure}

More importantly, the $R3$ phase belongs to a chiral space group in which both enantiomorphs are described by the same Sohncke space group~\cite{Bousquet_2025}. This implies that chirality and ferroelectricity coexist in this compound. We therefore investigate the correlation between left- and right-handedness and the two polarization states by computing chirality-related optical properties. Natural optical activity (NOA) is a specific gyrotropic effect manifested as a reciprocal rotation of the polarization plane of light in the absence of external magnetic fields, originating from the first-order spatial dispersion of the dielectric tensor~\cite{landau2013course,PhysRevLett.131.086902}. NOA is quantified by the gyration tensor $g_{\alpha\beta}$~\cite{PhysRevLett.131.086902,luo2025strainin}. Accordingly, we calculate the gyration tensor of $R3$ AgNbO$_3$ as a function of the polar distortion amplitude ($Q_{\Gamma_4^-}$). The $R3$ symmetry allows three independent gyrotropic tensor elements: $g_{11}=g_{22}$, $g_{33}$, and $g_{12}=-g_{21}$. Among them, $g_{12}$ exhibits the largest magnitude, as shown in Fig.~\ref{fig:fig2}(d), while the remaining components are presented in Fig.~\textcolor{magenta}{S12}. Upon varying $Q_{\Gamma_4^-}$, all tensor elements display pronounced nonlinear behavior, changing not only in magnitude but also in sign. The largest response is observed at $Q_{\Gamma_4^-}=\pm 0.2$~\AA, though in the R3 ground state $Q_{\Gamma_4^-} = \pm 0.43$~\AA, whereas all components vanish simultaneously with the disappearance of the $Q_{\Gamma_4^-}$ in the centrosymmetric $Im\bar{3}$ phase. Similarly, the second-order NLO coefficient $d_{16}$ shows a strong dependence on $Q_{\Gamma_4^-}$, exhibiting both amplitude modulation and sign reversal.

The circular dichroism (CD) spectrum normally exhibits a strong dependence on chiral handedness. For a pair of enantiomorphs, the CD spectra are expected to be equal in magnitude and opposite in sign, which is a hallmark of chiral materials. Figure~\ref{fig:fig4}(a) shows the calculated CD spectra of $R3$ AgNbO$_3$ as a function of $Q_{\Gamma_4^-}$. Indeed, the ferroelectric phase with positive polarization (red solid line) exhibits a CD peak at approximately 3.1~eV, while the phase with negative polarization (red dashed line) shows a peak at the same energy but with the opposite sign. As evident from Fig.~\ref{fig:fig4}(a), the position of the first CD peak depends sensitively on $Q_{\Gamma_4^-}$. Moreover, the sign of the CD spectrum is strictly determined by the polarization direction: all pairs of structures with opposite chiral handedness display identical magnitudes but opposite signs of CD over the entire energy range. Since CD is proportional to the Berry curvature~\cite{PhysRevB.77.235406}, we further plot the momentum-resolved Berry curvature of left- and right-handed AgNbO$_3$ in Figs.~\ref{fig:fig3}(c) and (d). The Berry curvature exhibits opposite signs at the same $\mathbf{k}$ points for the two polarization states, providing a microscopic explanation for the locking between polarization and the CD response in $R3$ AgNbO$_3$.

Circularly polarized light can also induce a direct-current (DC) photocurrent in non-centrosymmetric materials, known as the circular photogalvanic effect (CPGE). Although chirality is not a prerequisite for CPGE, chiral structures can host CPGE because they lack both inversion and mirror symmetries. Experimentally, CPGE is widely used to probe spin splitting, the nature of spin--orbit coupling, and nontrivial band topology~\cite{https://doi.org/10.1002/adom.202501190}. Under circularly polarized light excitation, a DC photocurrent is generated in chiral solids, providing information complementary to that obtained from CD measurements. Figure~\ref{fig:fig4}(b) presents the calculated CPGE response of $R3$ AgNbO$_3$ for different values of $Q_{\Gamma_4^-}$. For the structure with $Q_{\Gamma_4^-}=0.43$~\AA, the first CPGE peak appears at a photon energy of $\hbar\omega=2.8$~eV, followed by much stronger peaks at $\hbar\omega=3.2$ and 3.4~eV. As $Q_{\Gamma_4^-}$ decreases, the CPGE peaks shift toward lower photon energies, consistent with the reduction of the band gap E$_{\mathrm{g}}$, see Fig.~\ref{fig:fig2}(c). Notably, the CPGE intensity varies only weakly even when $Q_{\Gamma_4^-}$ is reduced to 42\% (0.18~\AA) of its original value. In contrast, the sign of the CPGE is directly correlated with the sign of $Q_{\Gamma_4^-}$, providing an electrical means to detect and control chirality. These results demonstrate that the chiral and nonlinear optical responses of $R3$ AgNbO$_3$ can be tuned by an external electric field, offering a promising route for electrically controllable optical information processing in optoelectronic applications.

In summary, through a systematic structural search based on symmetry-adapted phonon mode theory, we identify a previously unrecognized chiral polar structure with space group $R3$ in the complex perovskite AgNbO$_3$, which is energetically degenerate with the known lowest-energy phases. We further demonstrate that this polar phase is electrically switchable, owing to its large spontaneous polarization ($\sim$50~$\mu$C/cm$^{2}$) and a small switching energy barrier (5~meV/atom). More importantly, the $R3$ phase exhibits an intrinsic coupling between ferroelectric polarization and structural chirality, enabling electrical tuning of the chiral handedness and, consequently, of associated optical responses, including circular dichroism, circular photogalvanic effect, second harmonic generation, and natural optical activity. These results not only advance our understanding of the complex structural landscape of the long-studied AgNbO$_3$ compound but also identify a promising purely inorganic candidate for ultrafast, electrically controllable chiral photoelectronic devices.

\section{Acknowledgments}
Y.S, L.C., J.Z., and J.H. acknowledge the support received from the National Natural Science Foundation of China (Grant No.\,12374024). Y.Y. acknowledges the support received the support from the National Natural Science Foundation of China (Grant No.\,12304115), Fundamental Research Funds for the Central Universities (Grant No. FRF-TP-24-039A), and 2023 Fund for Fostering Young Scholars of the School of Mathematics and Physics, USTB (Grant No. FRF-BR-23-01B). Z.Y. acknowledges the support received from the National Natural Science Foundation of China (Grant No.\,12504046). L.B. is thankful for support from the Vannevar Bush Faculty Fellowship Grant No. N00014-20-1C2834 from the Department of Defense and an Impact Grant 3.0 from ARA.


\bibliography{ref}

\begin{thebibliography}{58}%
\makeatletter
\providecommand \@ifxundefined [1]{%
 \@ifx{#1\undefined}
}%
\providecommand \@ifnum [1]{%
 \ifnum #1\expandafter \@firstoftwo
 \else \expandafter \@secondoftwo
 \fi
}%
\providecommand \@ifx [1]{%
 \ifx #1\expandafter \@firstoftwo
 \else \expandafter \@secondoftwo
 \fi
}%
\providecommand \natexlab [1]{#1}%
\providecommand \enquote  [1]{``#1''}%
\providecommand \bibnamefont  [1]{#1}%
\providecommand \bibfnamefont [1]{#1}%
\providecommand \citenamefont [1]{#1}%
\providecommand \href@noop [0]{\@secondoftwo}%
\providecommand \href [0]{\begingroup \@sanitize@url \@href}%
\providecommand \@href[1]{\@@startlink{#1}\@@href}%
\providecommand \@@href[1]{\endgroup#1\@@endlink}%
\providecommand \@sanitize@url [0]{\catcode `\\12\catcode `\$12\catcode
  `\&12\catcode `\#12\catcode `\^12\catcode `\_12\catcode `\%12\relax}%
\providecommand \@@startlink[1]{}%
\providecommand \@@endlink[0]{}%
\providecommand \url  [0]{\begingroup\@sanitize@url \@url }%
\providecommand \@url [1]{\endgroup\@href {#1}{\urlprefix }}%
\providecommand \urlprefix  [0]{URL }%
\providecommand \Eprint [0]{\href }%
\providecommand \doibase [0]{http://dx.doi.org/}%
\providecommand \selectlanguage [0]{\@gobble}%
\providecommand \bibinfo  [0]{\@secondoftwo}%
\providecommand \bibfield  [0]{\@secondoftwo}%
\providecommand \translation [1]{[#1]}%
\providecommand \BibitemOpen [0]{}%
\providecommand \bibitemStop [0]{}%
\providecommand \bibitemNoStop [0]{.\EOS\space}%
\providecommand \EOS [0]{\spacefactor3000\relax}%
\providecommand \BibitemShut  [1]{\csname bibitem#1\endcsname}%
\let\auto@bib@innerbib\@empty
\bibitem [{\citenamefont {Guo}\ \emph {et~al.}(2025)\citenamefont {Guo},
  \citenamefont {Tian}, \citenamefont {Ma}, \citenamefont {Xu}, \citenamefont
  {Yang}, \citenamefont {She}, \citenamefont {Sun}, \citenamefont {Wu},
  \citenamefont {Ge}, \citenamefont {Jin} \emph {et~al.}}]{guo2025synergistic}%
  \BibitemOpen
  \bibfield  {author} {\bibinfo {author} {\bibfnamefont {S.}~\bibnamefont
  {Guo}}, \bibinfo {author} {\bibfnamefont {Y.}~\bibnamefont {Tian}}, \bibinfo
  {author} {\bibfnamefont {Q.}~\bibnamefont {Ma}}, \bibinfo {author}
  {\bibfnamefont {Y.}~\bibnamefont {Xu}}, \bibinfo {author} {\bibfnamefont
  {D.}~\bibnamefont {Yang}}, \bibinfo {author} {\bibfnamefont {L.}~\bibnamefont
  {She}}, \bibinfo {author} {\bibfnamefont {Z.}~\bibnamefont {Sun}}, \bibinfo
  {author} {\bibfnamefont {Y.}~\bibnamefont {Wu}}, \bibinfo {author}
  {\bibfnamefont {W.}~\bibnamefont {Ge}}, \bibinfo {author} {\bibfnamefont
  {L.}~\bibnamefont {Jin}},  \emph {et~al.},\ }\href@noop {} {\bibfield
  {journal} {\bibinfo  {journal} {Journal of the European Ceramic Society}\ ,\
  \bibinfo {pages} {117701}} (\bibinfo {year} {2025})}\BibitemShut {NoStop}%
\bibitem [{\citenamefont {Yang}\ \emph
  {et~al.}(2020{\natexlab{a}})\citenamefont {Yang}, \citenamefont {Gao},
  \citenamefont {Shu}, \citenamefont {Liu}, \citenamefont {Yu}, \citenamefont
  {Zhang}, \citenamefont {Wang}, \citenamefont {Zhang},\ and\ \citenamefont
  {Li}}]{yang2020lead}%
  \BibitemOpen
  \bibfield  {author} {\bibinfo {author} {\bibfnamefont {D.}~\bibnamefont
  {Yang}}, \bibinfo {author} {\bibfnamefont {J.}~\bibnamefont {Gao}}, \bibinfo
  {author} {\bibfnamefont {L.}~\bibnamefont {Shu}}, \bibinfo {author}
  {\bibfnamefont {Y.-X.}\ \bibnamefont {Liu}}, \bibinfo {author} {\bibfnamefont
  {J.}~\bibnamefont {Yu}}, \bibinfo {author} {\bibfnamefont {Y.}~\bibnamefont
  {Zhang}}, \bibinfo {author} {\bibfnamefont {X.}~\bibnamefont {Wang}},
  \bibinfo {author} {\bibfnamefont {B.-P.}\ \bibnamefont {Zhang}}, \ and\
  \bibinfo {author} {\bibfnamefont {J.-F.}\ \bibnamefont {Li}},\ }\href@noop {}
  {\bibfield  {journal} {\bibinfo  {journal} {Journal of Materials Chemistry
  A}\ }\textbf {\bibinfo {volume} {8}},\ \bibinfo {pages} {23724} (\bibinfo
  {year} {2020}{\natexlab{a}})}\BibitemShut {NoStop}%
\bibitem [{\citenamefont {Ting-Fang}\ \emph {et~al.}(2023)\citenamefont
  {Ting-Fang}, \citenamefont {Fei-Long}, \citenamefont {Chun-Chun},
  \citenamefont {Long-Long} \emph {et~al.}}]{ting2023advances}%
  \BibitemOpen
  \bibfield  {author} {\bibinfo {author} {\bibfnamefont {T.}~\bibnamefont
  {Ting-Fang}}, \bibinfo {author} {\bibfnamefont {C.}~\bibnamefont {Fei-Long}},
  \bibinfo {author} {\bibfnamefont {L.}~\bibnamefont {Chun-Chun}}, \bibinfo
  {author} {\bibfnamefont {S.}~\bibnamefont {Long-Long}},  \emph {et~al.},\
  }\href@noop {} {\bibfield  {journal} {\bibinfo  {journal} {Advanced
  Ceramics}\ }\textbf {\bibinfo {volume} {44}},\ \bibinfo {pages} {153}
  (\bibinfo {year} {2023})}\BibitemShut {NoStop}%
\bibitem [{\citenamefont {Xue}\ \emph {et~al.}(2025)\citenamefont {Xue},
  \citenamefont {Liu}, \citenamefont {Yang}, \citenamefont {Zhu}, \citenamefont
  {Yuan}, \citenamefont {Cao}, \citenamefont {Xiao}, \citenamefont {Chen},\
  and\ \citenamefont {Wang}}]{xue2025superior}%
  \BibitemOpen
  \bibfield  {author} {\bibinfo {author} {\bibfnamefont {R.}~\bibnamefont
  {Xue}}, \bibinfo {author} {\bibfnamefont {X.}~\bibnamefont {Liu}}, \bibinfo
  {author} {\bibfnamefont {Y.}~\bibnamefont {Yang}}, \bibinfo {author}
  {\bibfnamefont {X.}~\bibnamefont {Zhu}}, \bibinfo {author} {\bibfnamefont
  {C.}~\bibnamefont {Yuan}}, \bibinfo {author} {\bibfnamefont {B.}~\bibnamefont
  {Cao}}, \bibinfo {author} {\bibfnamefont {Y.}~\bibnamefont {Xiao}}, \bibinfo
  {author} {\bibfnamefont {Z.}~\bibnamefont {Chen}}, \ and\ \bibinfo {author}
  {\bibfnamefont {H.}~\bibnamefont {Wang}},\ }\href@noop {} {\bibfield
  {journal} {\bibinfo  {journal} {Ceramics International}\ } (\bibinfo {year}
  {2025})}\BibitemShut {NoStop}%
\bibitem [{\citenamefont {Zhang}\ \emph {et~al.}(2021)\citenamefont {Zhang},
  \citenamefont {Li}, \citenamefont {Song}, \citenamefont {Zhang},
  \citenamefont {Wang}, \citenamefont {Dai}, \citenamefont {Liu}, \citenamefont
  {Dong},\ and\ \citenamefont {Zhao}}]{zhang2021agnbo3}%
  \BibitemOpen
  \bibfield  {author} {\bibinfo {author} {\bibfnamefont {Y.}~\bibnamefont
  {Zhang}}, \bibinfo {author} {\bibfnamefont {X.}~\bibnamefont {Li}}, \bibinfo
  {author} {\bibfnamefont {J.}~\bibnamefont {Song}}, \bibinfo {author}
  {\bibfnamefont {S.}~\bibnamefont {Zhang}}, \bibinfo {author} {\bibfnamefont
  {J.}~\bibnamefont {Wang}}, \bibinfo {author} {\bibfnamefont {X.}~\bibnamefont
  {Dai}}, \bibinfo {author} {\bibfnamefont {B.}~\bibnamefont {Liu}}, \bibinfo
  {author} {\bibfnamefont {G.}~\bibnamefont {Dong}}, \ and\ \bibinfo {author}
  {\bibfnamefont {L.}~\bibnamefont {Zhao}},\ }\href@noop {} {\bibfield
  {journal} {\bibinfo  {journal} {Journal of Materiomics}\ }\textbf {\bibinfo
  {volume} {7}},\ \bibinfo {pages} {1294} (\bibinfo {year} {2021})}\BibitemShut
  {NoStop}%
\bibitem [{\citenamefont {Ma}\ \emph {et~al.}(2022)\citenamefont {Ma},
  \citenamefont {Li}, \citenamefont {Zhang}, \citenamefont {Duo}, \citenamefont
  {Zhang},\ and\ \citenamefont {Zhao}}]{ma2022dielectric}%
  \BibitemOpen
  \bibfield  {author} {\bibinfo {author} {\bibfnamefont {Q.}~\bibnamefont
  {Ma}}, \bibinfo {author} {\bibfnamefont {X.}~\bibnamefont {Li}}, \bibinfo
  {author} {\bibfnamefont {Y.}~\bibnamefont {Zhang}}, \bibinfo {author}
  {\bibfnamefont {Z.}~\bibnamefont {Duo}}, \bibinfo {author} {\bibfnamefont
  {S.}~\bibnamefont {Zhang}}, \ and\ \bibinfo {author} {\bibfnamefont
  {L.}~\bibnamefont {Zhao}},\ }\href@noop {} {\bibfield  {journal} {\bibinfo
  {journal} {Coatings}\ }\textbf {\bibinfo {volume} {12}},\ \bibinfo {pages}
  {1826} (\bibinfo {year} {2022})}\BibitemShut {NoStop}%
\bibitem [{\citenamefont {Tian}\ \emph {et~al.}(2022)\citenamefont {Tian},
  \citenamefont {Song}, \citenamefont {Viola}, \citenamefont {Shi},
  \citenamefont {Li}, \citenamefont {Jin}, \citenamefont {Hu}, \citenamefont
  {Xu}, \citenamefont {Ge}, \citenamefont {Yan} \emph
  {et~al.}}]{tian2022silver}%
  \BibitemOpen
  \bibfield  {author} {\bibinfo {author} {\bibfnamefont {Y.}~\bibnamefont
  {Tian}}, \bibinfo {author} {\bibfnamefont {P.}~\bibnamefont {Song}}, \bibinfo
  {author} {\bibfnamefont {G.}~\bibnamefont {Viola}}, \bibinfo {author}
  {\bibfnamefont {J.}~\bibnamefont {Shi}}, \bibinfo {author} {\bibfnamefont
  {J.}~\bibnamefont {Li}}, \bibinfo {author} {\bibfnamefont {L.}~\bibnamefont
  {Jin}}, \bibinfo {author} {\bibfnamefont {Q.}~\bibnamefont {Hu}}, \bibinfo
  {author} {\bibfnamefont {Y.}~\bibnamefont {Xu}}, \bibinfo {author}
  {\bibfnamefont {W.}~\bibnamefont {Ge}}, \bibinfo {author} {\bibfnamefont
  {Z.}~\bibnamefont {Yan}},  \emph {et~al.},\ }\href@noop {} {\bibfield
  {journal} {\bibinfo  {journal} {Journal of Materials Chemistry A}\ }\textbf
  {\bibinfo {volume} {10}},\ \bibinfo {pages} {14747} (\bibinfo {year}
  {2022})}\BibitemShut {NoStop}%
\bibitem [{\citenamefont {Yang}\ \emph
  {et~al.}(2020{\natexlab{b}})\citenamefont {Yang}, \citenamefont {Gao},
  \citenamefont {Shu}, \citenamefont {Liu}, \citenamefont {Yu}, \citenamefont
  {Zhang}, \citenamefont {Wang}, \citenamefont {Zhang},\ and\ \citenamefont
  {Li}}]{D0TA08345C}%
  \BibitemOpen
  \bibfield  {author} {\bibinfo {author} {\bibfnamefont {D.}~\bibnamefont
  {Yang}}, \bibinfo {author} {\bibfnamefont {J.}~\bibnamefont {Gao}}, \bibinfo
  {author} {\bibfnamefont {L.}~\bibnamefont {Shu}}, \bibinfo {author}
  {\bibfnamefont {Y.-X.}\ \bibnamefont {Liu}}, \bibinfo {author} {\bibfnamefont
  {J.}~\bibnamefont {Yu}}, \bibinfo {author} {\bibfnamefont {Y.}~\bibnamefont
  {Zhang}}, \bibinfo {author} {\bibfnamefont {X.}~\bibnamefont {Wang}},
  \bibinfo {author} {\bibfnamefont {B.-P.}\ \bibnamefont {Zhang}}, \ and\
  \bibinfo {author} {\bibfnamefont {J.-F.}\ \bibnamefont {Li}},\ }\href
  {\doibase 10.1039/D0TA08345C} {\bibfield  {journal} {\bibinfo  {journal} {J.
  Mater. Chem. A}\ }\textbf {\bibinfo {volume} {8}},\ \bibinfo {pages} {23724}
  (\bibinfo {year} {2020}{\natexlab{b}})}\BibitemShut {NoStop}%
\bibitem [{\citenamefont {Sciau}\ \emph
  {et~al.}(2004{\natexlab{a}})\citenamefont {Sciau}, \citenamefont {Kania},
  \citenamefont {Dkhil}, \citenamefont {Suard},\ and\ \citenamefont
  {Ratuszna}}]{sciau2004structural}%
  \BibitemOpen
  \bibfield  {author} {\bibinfo {author} {\bibfnamefont {P.}~\bibnamefont
  {Sciau}}, \bibinfo {author} {\bibfnamefont {A.}~\bibnamefont {Kania}},
  \bibinfo {author} {\bibfnamefont {B.}~\bibnamefont {Dkhil}}, \bibinfo
  {author} {\bibfnamefont {E.}~\bibnamefont {Suard}}, \ and\ \bibinfo {author}
  {\bibfnamefont {A.}~\bibnamefont {Ratuszna}},\ }\href@noop {} {\bibfield
  {journal} {\bibinfo  {journal} {Journal of Physics: Condensed Matter}\
  }\textbf {\bibinfo {volume} {16}},\ \bibinfo {pages} {2795} (\bibinfo {year}
  {2004}{\natexlab{a}})}\BibitemShut {NoStop}%
\bibitem [{\citenamefont {Sciau}\ \emph
  {et~al.}(2004{\natexlab{b}})\citenamefont {Sciau}, \citenamefont {Kania},
  \citenamefont {Dkhil}, \citenamefont {Suard},\ and\ \citenamefont
  {Ratuszna}}]{Ph_Sciau_2004}%
  \BibitemOpen
  \bibfield  {author} {\bibinfo {author} {\bibfnamefont {P.}~\bibnamefont
  {Sciau}}, \bibinfo {author} {\bibfnamefont {A.}~\bibnamefont {Kania}},
  \bibinfo {author} {\bibfnamefont {B.}~\bibnamefont {Dkhil}}, \bibinfo
  {author} {\bibfnamefont {E.}~\bibnamefont {Suard}}, \ and\ \bibinfo {author}
  {\bibfnamefont {A.}~\bibnamefont {Ratuszna}},\ }\href {\doibase
  10.1088/0953-8984/16/16/004} {\bibfield  {journal} {\bibinfo  {journal}
  {Journal of Physics: Condensed Matter}\ }\textbf {\bibinfo {volume} {16}},\
  \bibinfo {pages} {2795} (\bibinfo {year} {2004}{\natexlab{b}})}\BibitemShut
  {NoStop}%
\bibitem [{\citenamefont {Yashima}\ \emph {et~al.}(2011)\citenamefont
  {Yashima}, \citenamefont {Matsuyama}, \citenamefont {Sano}, \citenamefont
  {Itoh}, \citenamefont {Tsuda},\ and\ \citenamefont
  {Fu}}]{doi:10.1021/cm103389q}%
  \BibitemOpen
  \bibfield  {author} {\bibinfo {author} {\bibfnamefont {M.}~\bibnamefont
  {Yashima}}, \bibinfo {author} {\bibfnamefont {S.}~\bibnamefont {Matsuyama}},
  \bibinfo {author} {\bibfnamefont {R.}~\bibnamefont {Sano}}, \bibinfo {author}
  {\bibfnamefont {M.}~\bibnamefont {Itoh}}, \bibinfo {author} {\bibfnamefont
  {K.}~\bibnamefont {Tsuda}}, \ and\ \bibinfo {author} {\bibfnamefont
  {D.}~\bibnamefont {Fu}},\ }\href {\doibase 10.1021/cm103389q} {\bibfield
  {journal} {\bibinfo  {journal} {Chemistry of Materials}\ }\textbf {\bibinfo
  {volume} {23}},\ \bibinfo {pages} {1643} (\bibinfo {year}
  {2011})}\BibitemShut {NoStop}%
\bibitem [{\citenamefont {Fu}\ \emph {et~al.}(2007)\citenamefont {Fu},
  \citenamefont {Endo}, \citenamefont {Taniguchi}, \citenamefont {Taniyama},\
  and\ \citenamefont {Itoh}}]{10.1063/1.2751136}%
  \BibitemOpen
  \bibfield  {author} {\bibinfo {author} {\bibfnamefont {D.}~\bibnamefont
  {Fu}}, \bibinfo {author} {\bibfnamefont {M.}~\bibnamefont {Endo}}, \bibinfo
  {author} {\bibfnamefont {H.}~\bibnamefont {Taniguchi}}, \bibinfo {author}
  {\bibfnamefont {T.}~\bibnamefont {Taniyama}}, \ and\ \bibinfo {author}
  {\bibfnamefont {M.}~\bibnamefont {Itoh}},\ }\href {\doibase
  10.1063/1.2751136} {\bibfield  {journal} {\bibinfo  {journal} {Applied
  Physics Letters}\ }\textbf {\bibinfo {volume} {90}},\ \bibinfo {pages}
  {252907} (\bibinfo {year} {2007})}\BibitemShut {NoStop}%
\bibitem [{\citenamefont {Moriwake}\ \emph {et~al.}(2012)\citenamefont
  {Moriwake}, \citenamefont {Fisher}, \citenamefont {Kuwabara},\ and\
  \citenamefont {Fu}}]{Moriwake_2012}%
  \BibitemOpen
  \bibfield  {author} {\bibinfo {author} {\bibfnamefont {H.}~\bibnamefont
  {Moriwake}}, \bibinfo {author} {\bibfnamefont {C.~A.~J.}\ \bibnamefont
  {Fisher}}, \bibinfo {author} {\bibfnamefont {A.}~\bibnamefont {Kuwabara}}, \
  and\ \bibinfo {author} {\bibfnamefont {D.}~\bibnamefont {Fu}},\ }\href
  {\doibase 10.1143/JJAP.51.09LE02} {\bibfield  {journal} {\bibinfo  {journal}
  {Japanese Journal of Applied Physics}\ }\textbf {\bibinfo {volume} {51}},\
  \bibinfo {pages} {09LE02} (\bibinfo {year} {2012})}\BibitemShut {NoStop}%
\bibitem [{\citenamefont {Tian}\ \emph {et~al.}(2016)\citenamefont {Tian},
  \citenamefont {Jin}, \citenamefont {Zhang}, \citenamefont {Xu}, \citenamefont
  {Wei}, \citenamefont {Politova}, \citenamefont {Stefanovich}, \citenamefont
  {Tarakina}, \citenamefont {Abrahams},\ and\ \citenamefont
  {Yan}}]{C6TA06353E}%
  \BibitemOpen
  \bibfield  {author} {\bibinfo {author} {\bibfnamefont {Y.}~\bibnamefont
  {Tian}}, \bibinfo {author} {\bibfnamefont {L.}~\bibnamefont {Jin}}, \bibinfo
  {author} {\bibfnamefont {H.}~\bibnamefont {Zhang}}, \bibinfo {author}
  {\bibfnamefont {Z.}~\bibnamefont {Xu}}, \bibinfo {author} {\bibfnamefont
  {X.}~\bibnamefont {Wei}}, \bibinfo {author} {\bibfnamefont {E.~D.}\
  \bibnamefont {Politova}}, \bibinfo {author} {\bibfnamefont {S.~Y.}\
  \bibnamefont {Stefanovich}}, \bibinfo {author} {\bibfnamefont {N.~V.}\
  \bibnamefont {Tarakina}}, \bibinfo {author} {\bibfnamefont {I.}~\bibnamefont
  {Abrahams}}, \ and\ \bibinfo {author} {\bibfnamefont {H.}~\bibnamefont
  {Yan}},\ }\href {\doibase 10.1039/C6TA06353E} {\bibfield  {journal} {\bibinfo
   {journal} {J. Mater. Chem. A}\ }\textbf {\bibinfo {volume} {4}},\ \bibinfo
  {pages} {17279} (\bibinfo {year} {2016})}\BibitemShut {NoStop}%
\bibitem [{\citenamefont {Shigemi}\ and\ \citenamefont
  {Wada}(2008)}]{AgNbO3-R3c}%
  \BibitemOpen
  \bibfield  {author} {\bibinfo {author} {\bibfnamefont {A.}~\bibnamefont
  {Shigemi}}\ and\ \bibinfo {author} {\bibfnamefont {T.}~\bibnamefont {Wada}},\
  }\href {\doibase 10.1080/08927020802235698} {\bibfield  {journal} {\bibinfo
  {journal} {Mol. Simul.}\ }\textbf {\bibinfo {volume} {34}},\ \bibinfo {pages}
  {1105} (\bibinfo {year} {2008})}\BibitemShut {NoStop}%
\bibitem [{\citenamefont {Zhang}\ \emph {et~al.}(2024)\citenamefont {Zhang},
  \citenamefont {Shapovalov}, \citenamefont {Amisi},\ and\ \citenamefont
  {Ghosez}}]{zhang2024lattice}%
  \BibitemOpen
  \bibfield  {author} {\bibinfo {author} {\bibfnamefont {H.}~\bibnamefont
  {Zhang}}, \bibinfo {author} {\bibfnamefont {K.}~\bibnamefont {Shapovalov}},
  \bibinfo {author} {\bibfnamefont {S.}~\bibnamefont {Amisi}}, \ and\ \bibinfo
  {author} {\bibfnamefont {P.}~\bibnamefont {Ghosez}},\ }\href@noop {}
  {\bibfield  {journal} {\bibinfo  {journal} {Physical Review B}\ }\textbf
  {\bibinfo {volume} {110}},\ \bibinfo {pages} {064305} (\bibinfo {year}
  {2024})}\BibitemShut {NoStop}%
\bibitem [{\citenamefont {Tatsuzaki}\ \emph {et~al.}(1966)\citenamefont
  {Tatsuzaki}, \citenamefont {Itoh}, \citenamefont {Ueda},\ and\ \citenamefont
  {Shindo}}]{PhysRevLett.17.198}%
  \BibitemOpen
  \bibfield  {author} {\bibinfo {author} {\bibfnamefont {I.}~\bibnamefont
  {Tatsuzaki}}, \bibinfo {author} {\bibfnamefont {K.}~\bibnamefont {Itoh}},
  \bibinfo {author} {\bibfnamefont {S.}~\bibnamefont {Ueda}}, \ and\ \bibinfo
  {author} {\bibfnamefont {Y.}~\bibnamefont {Shindo}},\ }\href {\doibase
  10.1103/PhysRevLett.17.198} {\bibfield  {journal} {\bibinfo  {journal} {Phys.
  Rev. Lett.}\ }\textbf {\bibinfo {volume} {17}},\ \bibinfo {pages} {198}
  (\bibinfo {year} {1966})}\BibitemShut {NoStop}%
\bibitem [{\citenamefont {Kundys}\ \emph {et~al.}(2010)\citenamefont {Kundys},
  \citenamefont {Viret}, \citenamefont {Colson},\ and\ \citenamefont
  {Kundys}}]{kundys2010light}%
  \BibitemOpen
  \bibfield  {author} {\bibinfo {author} {\bibfnamefont {B.}~\bibnamefont
  {Kundys}}, \bibinfo {author} {\bibfnamefont {M.}~\bibnamefont {Viret}},
  \bibinfo {author} {\bibfnamefont {D.}~\bibnamefont {Colson}}, \ and\ \bibinfo
  {author} {\bibfnamefont {D.~O.}\ \bibnamefont {Kundys}},\ }\href {\doibase
  doi.org/10.1038/nmat2807} {\bibfield  {journal} {\bibinfo  {journal} {Nat.
  Mater.}\ }\textbf {\bibinfo {volume} {9}},\ \bibinfo {pages} {803} (\bibinfo
  {year} {2010})}\BibitemShut {NoStop}%
\bibitem [{\citenamefont {Kundys}(2015)}]{10.1063/1.4905505}%
  \BibitemOpen
  \bibfield  {author} {\bibinfo {author} {\bibfnamefont {B.}~\bibnamefont
  {Kundys}},\ }\href {\doibase 10.1063/1.4905505} {\bibfield  {journal}
  {\bibinfo  {journal} {Appl. Phys. Rev.}\ }\textbf {\bibinfo {volume} {2}},\
  \bibinfo {pages} {011301} (\bibinfo {year} {2015})}\BibitemShut {NoStop}%
\bibitem [{\citenamefont {Li}\ \emph {et~al.}(2024)\citenamefont {Li},
  \citenamefont {Varrassi}, \citenamefont {Yang}, \citenamefont {Franchini},
  \citenamefont {Bellaiche},\ and\ \citenamefont
  {He}}]{doi:10.1021/jacs.4c03296}%
  \BibitemOpen
  \bibfield  {author} {\bibinfo {author} {\bibfnamefont {Z.}~\bibnamefont
  {Li}}, \bibinfo {author} {\bibfnamefont {L.}~\bibnamefont {Varrassi}},
  \bibinfo {author} {\bibfnamefont {Y.}~\bibnamefont {Yang}}, \bibinfo {author}
  {\bibfnamefont {C.}~\bibnamefont {Franchini}}, \bibinfo {author}
  {\bibfnamefont {L.}~\bibnamefont {Bellaiche}}, \ and\ \bibinfo {author}
  {\bibfnamefont {J.}~\bibnamefont {He}},\ }\href {\doibase
  10.1021/jacs.4c03296} {\bibfield  {journal} {\bibinfo  {journal} {Journal of
  the American Chemical Society}\ }\textbf {\bibinfo {volume} {146}},\ \bibinfo
  {pages} {15411} (\bibinfo {year} {2024})},\ \bibinfo {note} {pMID:
  38780106}\BibitemShut {NoStop}%
\bibitem [{\citenamefont {He}\ \emph {et~al.}(2025)\citenamefont {He},
  \citenamefont {Chen}, \citenamefont {Wu}, \citenamefont {Wong}, \citenamefont
  {Wu}, \citenamefont {Chang}, \citenamefont {Wang}, \citenamefont {Gao},\ and\
  \citenamefont {Loh}}]{doi:10.1021/jacs.4c13604}%
  \BibitemOpen
  \bibfield  {author} {\bibinfo {author} {\bibfnamefont {W.}~\bibnamefont
  {He}}, \bibinfo {author} {\bibfnamefont {C.}~\bibnamefont {Chen}}, \bibinfo
  {author} {\bibfnamefont {S.}~\bibnamefont {Wu}}, \bibinfo {author}
  {\bibfnamefont {W.~P.~D.}\ \bibnamefont {Wong}}, \bibinfo {author}
  {\bibfnamefont {Z.}~\bibnamefont {Wu}}, \bibinfo {author} {\bibfnamefont
  {K.}~\bibnamefont {Chang}}, \bibinfo {author} {\bibfnamefont
  {J.}~\bibnamefont {Wang}}, \bibinfo {author} {\bibfnamefont {H.}~\bibnamefont
  {Gao}}, \ and\ \bibinfo {author} {\bibfnamefont {K.~P.}\ \bibnamefont
  {Loh}},\ }\href {\doibase 10.1021/jacs.4c13604} {\bibfield  {journal}
  {\bibinfo  {journal} {Journal of the American Chemical Society}\ }\textbf
  {\bibinfo {volume} {147}},\ \bibinfo {pages} {811} (\bibinfo {year}
  {2025})},\ \bibinfo {note} {pMID: 39699582}\BibitemShut {NoStop}%
\bibitem [{\citenamefont {Fiebig}\ \emph {et~al.}(2016)\citenamefont {Fiebig},
  \citenamefont {Lottermoser}, \citenamefont {Meier},\ and\ \citenamefont
  {Trassin}}]{fiebig2016evolution}%
  \BibitemOpen
  \bibfield  {author} {\bibinfo {author} {\bibfnamefont {M.}~\bibnamefont
  {Fiebig}}, \bibinfo {author} {\bibfnamefont {T.}~\bibnamefont {Lottermoser}},
  \bibinfo {author} {\bibfnamefont {D.}~\bibnamefont {Meier}}, \ and\ \bibinfo
  {author} {\bibfnamefont {M.}~\bibnamefont {Trassin}},\ }\href@noop {}
  {\bibfield  {journal} {\bibinfo  {journal} {Nature Reviews Materials}\
  }\textbf {\bibinfo {volume} {1}},\ \bibinfo {pages} {1} (\bibinfo {year}
  {2016})}\BibitemShut {NoStop}%
\bibitem [{\citenamefont {Huang}\ \emph {et~al.}(2024)\citenamefont {Huang},
  \citenamefont {Chen}, \citenamefont {Li}, \citenamefont {Mangeri},
  \citenamefont {Zhang}, \citenamefont {Ramesh}, \citenamefont {Taghinejad},
  \citenamefont {Meisenheimer}, \citenamefont {Caretta}, \citenamefont
  {Susarla} \emph {et~al.}}]{huang2024manipulating}%
  \BibitemOpen
  \bibfield  {author} {\bibinfo {author} {\bibfnamefont {X.}~\bibnamefont
  {Huang}}, \bibinfo {author} {\bibfnamefont {X.}~\bibnamefont {Chen}},
  \bibinfo {author} {\bibfnamefont {Y.}~\bibnamefont {Li}}, \bibinfo {author}
  {\bibfnamefont {J.}~\bibnamefont {Mangeri}}, \bibinfo {author} {\bibfnamefont
  {H.}~\bibnamefont {Zhang}}, \bibinfo {author} {\bibfnamefont
  {M.}~\bibnamefont {Ramesh}}, \bibinfo {author} {\bibfnamefont
  {H.}~\bibnamefont {Taghinejad}}, \bibinfo {author} {\bibfnamefont
  {P.}~\bibnamefont {Meisenheimer}}, \bibinfo {author} {\bibfnamefont
  {L.}~\bibnamefont {Caretta}}, \bibinfo {author} {\bibfnamefont
  {S.}~\bibnamefont {Susarla}},  \emph {et~al.},\ }\href@noop {} {\bibfield
  {journal} {\bibinfo  {journal} {Nature materials}\ }\textbf {\bibinfo
  {volume} {23}},\ \bibinfo {pages} {898} (\bibinfo {year} {2024})}\BibitemShut
  {NoStop}%
\bibitem [{\citenamefont {Di~Sante}\ \emph {et~al.}(2013)\citenamefont
  {Di~Sante}, \citenamefont {Barone}, \citenamefont {Bertacco},\ and\
  \citenamefont {Picozzi}}]{https://doi.org/10.1002/adma.201203199}%
  \BibitemOpen
  \bibfield  {author} {\bibinfo {author} {\bibfnamefont {D.}~\bibnamefont
  {Di~Sante}}, \bibinfo {author} {\bibfnamefont {P.}~\bibnamefont {Barone}},
  \bibinfo {author} {\bibfnamefont {R.}~\bibnamefont {Bertacco}}, \ and\
  \bibinfo {author} {\bibfnamefont {S.}~\bibnamefont {Picozzi}},\ }\href
  {\doibase https://doi.org/10.1002/adma.201203199} {\bibfield  {journal}
  {\bibinfo  {journal} {Adv. Mater.}\ }\textbf {\bibinfo {volume} {25}},\
  \bibinfo {pages} {509} (\bibinfo {year} {2013})}\BibitemShut {NoStop}%
\bibitem [{\citenamefont {Rinaldi}\ \emph {et~al.}(2018)\citenamefont
  {Rinaldi}, \citenamefont {Varotto}, \citenamefont {Asa}, \citenamefont
  {Sławińska}, \citenamefont {Fujii}, \citenamefont {Vinai}, \citenamefont
  {Cecchi}, \citenamefont {Di~Sante}, \citenamefont {Calarco}, \citenamefont
  {Vobornik}, \citenamefont {Panaccione}, \citenamefont {Picozzi},\ and\
  \citenamefont {Bertacco}}]{doi:10.1021/acs.nanolett.7b04829}%
  \BibitemOpen
  \bibfield  {author} {\bibinfo {author} {\bibfnamefont {C.}~\bibnamefont
  {Rinaldi}}, \bibinfo {author} {\bibfnamefont {S.}~\bibnamefont {Varotto}},
  \bibinfo {author} {\bibfnamefont {M.}~\bibnamefont {Asa}}, \bibinfo {author}
  {\bibfnamefont {J.}~\bibnamefont {Sławińska}}, \bibinfo {author}
  {\bibfnamefont {J.}~\bibnamefont {Fujii}}, \bibinfo {author} {\bibfnamefont
  {G.}~\bibnamefont {Vinai}}, \bibinfo {author} {\bibfnamefont
  {S.}~\bibnamefont {Cecchi}}, \bibinfo {author} {\bibfnamefont
  {D.}~\bibnamefont {Di~Sante}}, \bibinfo {author} {\bibfnamefont
  {R.}~\bibnamefont {Calarco}}, \bibinfo {author} {\bibfnamefont
  {I.}~\bibnamefont {Vobornik}}, \bibinfo {author} {\bibfnamefont
  {G.}~\bibnamefont {Panaccione}}, \bibinfo {author} {\bibfnamefont
  {S.}~\bibnamefont {Picozzi}}, \ and\ \bibinfo {author} {\bibfnamefont
  {R.}~\bibnamefont {Bertacco}},\ }\href {\doibase
  10.1021/acs.nanolett.7b04829} {\bibfield  {journal} {\bibinfo  {journal}
  {Nano Lett.}\ }\textbf {\bibinfo {volume} {18}},\ \bibinfo {pages} {2751}
  (\bibinfo {year} {2018})}\BibitemShut {NoStop}%
\bibitem [{\citenamefont {Liebmann}\ \emph {et~al.}(2016)\citenamefont
  {Liebmann}, \citenamefont {Rinaldi}, \citenamefont {Di~Sante}, \citenamefont
  {Kellner}, \citenamefont {Pauly}, \citenamefont {Wang}, \citenamefont
  {Boschker}, \citenamefont {Giussani}, \citenamefont {Bertoli}, \citenamefont
  {Cantoni} \emph {et~al.}}]{liebmann2016giant}%
  \BibitemOpen
  \bibfield  {author} {\bibinfo {author} {\bibfnamefont {M.}~\bibnamefont
  {Liebmann}}, \bibinfo {author} {\bibfnamefont {C.}~\bibnamefont {Rinaldi}},
  \bibinfo {author} {\bibfnamefont {D.}~\bibnamefont {Di~Sante}}, \bibinfo
  {author} {\bibfnamefont {J.}~\bibnamefont {Kellner}}, \bibinfo {author}
  {\bibfnamefont {C.}~\bibnamefont {Pauly}}, \bibinfo {author} {\bibfnamefont
  {R.~N.}\ \bibnamefont {Wang}}, \bibinfo {author} {\bibfnamefont {J.~E.}\
  \bibnamefont {Boschker}}, \bibinfo {author} {\bibfnamefont {A.}~\bibnamefont
  {Giussani}}, \bibinfo {author} {\bibfnamefont {S.}~\bibnamefont {Bertoli}},
  \bibinfo {author} {\bibfnamefont {M.}~\bibnamefont {Cantoni}},  \emph
  {et~al.},\ }\href {\doibase doi.org/10.1002/adma.201503459} {\bibfield
  {journal} {\bibinfo  {journal} {Adv. Mater.}\ }\textbf {\bibinfo {volume}
  {28}},\ \bibinfo {pages} {560} (\bibinfo {year} {2016})}\BibitemShut
  {NoStop}%
\bibitem [{\citenamefont {He}\ \emph {et~al.}(2018)\citenamefont {He},
  \citenamefont {Di~Sante}, \citenamefont {Li}, \citenamefont {Chen},
  \citenamefont {Rondinelli},\ and\ \citenamefont {Franchini}}]{he2018tunable}%
  \BibitemOpen
  \bibfield  {author} {\bibinfo {author} {\bibfnamefont {J.}~\bibnamefont
  {He}}, \bibinfo {author} {\bibfnamefont {D.}~\bibnamefont {Di~Sante}},
  \bibinfo {author} {\bibfnamefont {R.}~\bibnamefont {Li}}, \bibinfo {author}
  {\bibfnamefont {X.-Q.}\ \bibnamefont {Chen}}, \bibinfo {author}
  {\bibfnamefont {J.~M.}\ \bibnamefont {Rondinelli}}, \ and\ \bibinfo {author}
  {\bibfnamefont {C.}~\bibnamefont {Franchini}},\ }\href@noop {} {\bibfield
  {journal} {\bibinfo  {journal} {Nature communications}\ }\textbf {\bibinfo
  {volume} {9}},\ \bibinfo {pages} {492} (\bibinfo {year} {2018})}\BibitemShut
  {NoStop}%
\bibitem [{\citenamefont {Duan}\ \emph {et~al.}(2025)\citenamefont {Duan},
  \citenamefont {Zhang}, \citenamefont {Zhu}, \citenamefont {Liu},
  \citenamefont {Zhang}, \citenamefont {\ifmmode \check{Z}\else
  \v{Z}\fi{}uti\ifmmode~\acute{c}\else \'{c}\fi{}},\ and\ \citenamefont
  {Zhou}}]{PhysRevLett.134.106801}%
  \BibitemOpen
  \bibfield  {author} {\bibinfo {author} {\bibfnamefont {X.}~\bibnamefont
  {Duan}}, \bibinfo {author} {\bibfnamefont {J.}~\bibnamefont {Zhang}},
  \bibinfo {author} {\bibfnamefont {Z.}~\bibnamefont {Zhu}}, \bibinfo {author}
  {\bibfnamefont {Y.}~\bibnamefont {Liu}}, \bibinfo {author} {\bibfnamefont
  {Z.}~\bibnamefont {Zhang}}, \bibinfo {author} {\bibfnamefont
  {I.}~\bibnamefont {\ifmmode \check{Z}\else
  \v{Z}\fi{}uti\ifmmode~\acute{c}\else \'{c}\fi{}}}, \ and\ \bibinfo {author}
  {\bibfnamefont {T.}~\bibnamefont {Zhou}},\ }\href {\doibase
  10.1103/PhysRevLett.134.106801} {\bibfield  {journal} {\bibinfo  {journal}
  {Phys. Rev. Lett.}\ }\textbf {\bibinfo {volume} {134}},\ \bibinfo {pages}
  {106801} (\bibinfo {year} {2025})}\BibitemShut {NoStop}%
\bibitem [{\citenamefont {Niu}\ \emph {et~al.}(2023)\citenamefont {Niu},
  \citenamefont {Zhao}, \citenamefont {Yan}, \citenamefont {Pang},
  \citenamefont {Li}, \citenamefont {Yang},\ and\ \citenamefont
  {Wang}}]{niu2023chiral}%
  \BibitemOpen
  \bibfield  {author} {\bibinfo {author} {\bibfnamefont {X.}~\bibnamefont
  {Niu}}, \bibinfo {author} {\bibfnamefont {R.}~\bibnamefont {Zhao}}, \bibinfo
  {author} {\bibfnamefont {S.}~\bibnamefont {Yan}}, \bibinfo {author}
  {\bibfnamefont {Z.}~\bibnamefont {Pang}}, \bibinfo {author} {\bibfnamefont
  {H.}~\bibnamefont {Li}}, \bibinfo {author} {\bibfnamefont {X.}~\bibnamefont
  {Yang}}, \ and\ \bibinfo {author} {\bibfnamefont {K.}~\bibnamefont {Wang}},\
  }\href@noop {} {\bibfield  {journal} {\bibinfo  {journal} {Small}\ }\textbf
  {\bibinfo {volume} {19}},\ \bibinfo {pages} {2303059} (\bibinfo {year}
  {2023})}\BibitemShut {NoStop}%
\bibitem [{\citenamefont {Bousquet}\ \emph {et~al.}(2025)\citenamefont
  {Bousquet}, \citenamefont {Fava}, \citenamefont {Romestan}, \citenamefont
  {Gómez-Ortiz}, \citenamefont {McCabe},\ and\ \citenamefont
  {Romero}}]{Bousquet_2025}%
  \BibitemOpen
  \bibfield  {author} {\bibinfo {author} {\bibfnamefont {E.}~\bibnamefont
  {Bousquet}}, \bibinfo {author} {\bibfnamefont {M.}~\bibnamefont {Fava}},
  \bibinfo {author} {\bibfnamefont {Z.}~\bibnamefont {Romestan}}, \bibinfo
  {author} {\bibfnamefont {F.}~\bibnamefont {Gómez-Ortiz}}, \bibinfo {author}
  {\bibfnamefont {E.~E.}\ \bibnamefont {McCabe}}, \ and\ \bibinfo {author}
  {\bibfnamefont {A.~H.}\ \bibnamefont {Romero}},\ }\href {\doibase
  10.1088/1361-648X/adb674} {\bibfield  {journal} {\bibinfo  {journal} {Journal
  of Physics: Condensed Matter}\ }\textbf {\bibinfo {volume} {37}},\ \bibinfo
  {pages} {163004} (\bibinfo {year} {2025})}\BibitemShut {NoStop}%
\bibitem [{\citenamefont {Powers}(2013)}]{powers2013field}%
  \BibitemOpen
  \bibfield  {author} {\bibinfo {author} {\bibfnamefont {P.~E.}\ \bibnamefont
  {Powers}},\ }\href@noop {} {\emph {\bibinfo {title} {Field guide to nonlinear
  optics}}}\ (\bibinfo  {publisher} {SPIE, Bellingham, Washington, USA},\
  \bibinfo {year} {2013})\BibitemShut {NoStop}%
\bibitem [{\citenamefont {Nussenzveig}(1973)}]{nussenzveig1973introduction}%
  \BibitemOpen
  \bibfield  {author} {\bibinfo {author} {\bibfnamefont {H.~M.}\ \bibnamefont
  {Nussenzveig}},\ }\href@noop {} {\emph {\bibinfo {title} {Introduction to
  quantum optics}}}\ (\bibinfo  {publisher} {CRC Press},\ \bibinfo {year}
  {1973})\BibitemShut {NoStop}%
\bibitem [{\citenamefont {Caldwell}(1966)}]{doi:10.1073/pnas.56.5.1391}%
  \BibitemOpen
  \bibfield  {author} {\bibinfo {author} {\bibfnamefont {D.~J.}\ \bibnamefont
  {Caldwell}},\ }\href {\doibase 10.1073/pnas.56.5.1391} {\bibfield  {journal}
  {\bibinfo  {journal} {Proceedings of the National Academy of Sciences}\
  }\textbf {\bibinfo {volume} {56}},\ \bibinfo {pages} {1391} (\bibinfo {year}
  {1966})},\ \Eprint
  {http://arxiv.org/abs/https://www.pnas.org/doi/pdf/10.1073/pnas.56.5.1391}
  {https://www.pnas.org/doi/pdf/10.1073/pnas.56.5.1391} \BibitemShut {NoStop}%
\bibitem [{\citenamefont {{\v{Z}}uti{\'c}}, \citenamefont {Fabian},\ and\
  \citenamefont {Sarma}(2004)}]{vzutic2004spintronics}%
  \BibitemOpen
  \bibfield  {author} {\bibinfo {author} {\bibfnamefont {I.}~\bibnamefont
  {{\v{Z}}uti{\'c}}}, \bibinfo {author} {\bibfnamefont {J.}~\bibnamefont
  {Fabian}}, \ and\ \bibinfo {author} {\bibfnamefont {S.~D.}\ \bibnamefont
  {Sarma}},\ }\href@noop {} {\bibfield  {journal} {\bibinfo  {journal} {Reviews
  of modern physics}\ }\textbf {\bibinfo {volume} {76}},\ \bibinfo {pages}
  {323} (\bibinfo {year} {2004})}\BibitemShut {NoStop}%
\bibitem [{\citenamefont {Le}\ and\ \citenamefont
  {Sun}(2021)}]{le2021topology}%
  \BibitemOpen
  \bibfield  {author} {\bibinfo {author} {\bibfnamefont {C.}~\bibnamefont
  {Le}}\ and\ \bibinfo {author} {\bibfnamefont {Y.}~\bibnamefont {Sun}},\
  }\href@noop {} {\bibfield  {journal} {\bibinfo  {journal} {Journal of
  Physics: Condensed Matter}\ }\textbf {\bibinfo {volume} {33}},\ \bibinfo
  {pages} {503003} (\bibinfo {year} {2021})}\BibitemShut {NoStop}%
\bibitem [{\citenamefont {De~Juan}\ \emph {et~al.}(2017)\citenamefont
  {De~Juan}, \citenamefont {Grushin}, \citenamefont {Morimoto},\ and\
  \citenamefont {Moore}}]{de2017quantized}%
  \BibitemOpen
  \bibfield  {author} {\bibinfo {author} {\bibfnamefont {F.}~\bibnamefont
  {De~Juan}}, \bibinfo {author} {\bibfnamefont {A.~G.}\ \bibnamefont
  {Grushin}}, \bibinfo {author} {\bibfnamefont {T.}~\bibnamefont {Morimoto}}, \
  and\ \bibinfo {author} {\bibfnamefont {J.~E.}\ \bibnamefont {Moore}},\
  }\href@noop {} {\bibfield  {journal} {\bibinfo  {journal} {Nature
  communications}\ }\textbf {\bibinfo {volume} {8}},\ \bibinfo {pages} {15995}
  (\bibinfo {year} {2017})}\BibitemShut {NoStop}%
\bibitem [{\citenamefont {Le}\ \emph {et~al.}(2020)\citenamefont {Le},
  \citenamefont {Zhang}, \citenamefont {Felser},\ and\ \citenamefont
  {Sun}}]{le2020ab}%
  \BibitemOpen
  \bibfield  {author} {\bibinfo {author} {\bibfnamefont {C.}~\bibnamefont
  {Le}}, \bibinfo {author} {\bibfnamefont {Y.}~\bibnamefont {Zhang}}, \bibinfo
  {author} {\bibfnamefont {C.}~\bibnamefont {Felser}}, \ and\ \bibinfo {author}
  {\bibfnamefont {Y.}~\bibnamefont {Sun}},\ }\href@noop {} {\bibfield
  {journal} {\bibinfo  {journal} {Physical Review B}\ }\textbf {\bibinfo
  {volume} {102}},\ \bibinfo {pages} {121111} (\bibinfo {year}
  {2020})}\BibitemShut {NoStop}%
\bibitem [{\citenamefont {Dong}\ \emph {et~al.}(2019)\citenamefont {Dong},
  \citenamefont {Zhang}, \citenamefont {Li}, \citenamefont {Feng},
  \citenamefont {Zhang},\ and\ \citenamefont
  {Xu}}]{https://doi.org/10.1002/smll.201902237}%
  \BibitemOpen
  \bibfield  {author} {\bibinfo {author} {\bibfnamefont {Y.}~\bibnamefont
  {Dong}}, \bibinfo {author} {\bibfnamefont {Y.}~\bibnamefont {Zhang}},
  \bibinfo {author} {\bibfnamefont {X.}~\bibnamefont {Li}}, \bibinfo {author}
  {\bibfnamefont {Y.}~\bibnamefont {Feng}}, \bibinfo {author} {\bibfnamefont
  {H.}~\bibnamefont {Zhang}}, \ and\ \bibinfo {author} {\bibfnamefont
  {J.}~\bibnamefont {Xu}},\ }\href {\doibase
  https://doi.org/10.1002/smll.201902237} {\bibfield  {journal} {\bibinfo
  {journal} {Small}\ }\textbf {\bibinfo {volume} {15}},\ \bibinfo {pages}
  {1902237} (\bibinfo {year} {2019})}\BibitemShut {NoStop}%
\bibitem [{\citenamefont {Dang}\ \emph {et~al.}(2021)\citenamefont {Dang},
  \citenamefont {Liu}, \citenamefont {Cao},\ and\ \citenamefont
  {Tao}}]{DANG2021794}%
  \BibitemOpen
  \bibfield  {author} {\bibinfo {author} {\bibfnamefont {Y.}~\bibnamefont
  {Dang}}, \bibinfo {author} {\bibfnamefont {X.}~\bibnamefont {Liu}}, \bibinfo
  {author} {\bibfnamefont {B.}~\bibnamefont {Cao}}, \ and\ \bibinfo {author}
  {\bibfnamefont {X.}~\bibnamefont {Tao}},\ }\href {\doibase
  https://doi.org/10.1016/j.matt.2020.12.018} {\bibfield  {journal} {\bibinfo
  {journal} {Matter}\ }\textbf {\bibinfo {volume} {4}},\ \bibinfo {pages} {794}
  (\bibinfo {year} {2021})}\BibitemShut {NoStop}%
\bibitem [{\citenamefont {Luo}\ \emph {et~al.}(2025)\citenamefont {Luo},
  \citenamefont {Zabalo}, \citenamefont {Ren}, \citenamefont {Jung},
  \citenamefont {Stengel}, \citenamefont {Mishra}, \citenamefont
  {Ravichandran},\ and\ \citenamefont {Bellaiche}}]{luo2025strainin}%
  \BibitemOpen
  \bibfield  {author} {\bibinfo {author} {\bibfnamefont {W.}~\bibnamefont
  {Luo}}, \bibinfo {author} {\bibfnamefont {A.}~\bibnamefont {Zabalo}},
  \bibinfo {author} {\bibfnamefont {G.}~\bibnamefont {Ren}}, \bibinfo {author}
  {\bibfnamefont {G.-Y.}\ \bibnamefont {Jung}}, \bibinfo {author}
  {\bibfnamefont {M.}~\bibnamefont {Stengel}}, \bibinfo {author} {\bibfnamefont
  {R.}~\bibnamefont {Mishra}}, \bibinfo {author} {\bibfnamefont
  {J.}~\bibnamefont {Ravichandran}}, \ and\ \bibinfo {author} {\bibfnamefont
  {L.}~\bibnamefont {Bellaiche}},\ }\href {https://arxiv.org/abs/2505.09881}
  {\enquote {\bibinfo {title} {Strain-induced gyrotropic effects in
  ferroelectric batis3},}\ } (\bibinfo {year} {2025}),\ \Eprint
  {http://arxiv.org/abs/2505.09881} {arXiv:2505.09881 [cond-mat.mtrl-sci]}
  \BibitemShut {NoStop}%
\bibitem [{\citenamefont {Gutierrez-Amigo}\ \emph {et~al.}(2025)\citenamefont
  {Gutierrez-Amigo}, \citenamefont {Felser}, \citenamefont {Errea},\ and\
  \citenamefont {Vergniory}}]{gutierrezamigo2025}%
  \BibitemOpen
  \bibfield  {author} {\bibinfo {author} {\bibfnamefont {M.}~\bibnamefont
  {Gutierrez-Amigo}}, \bibinfo {author} {\bibfnamefont {C.}~\bibnamefont
  {Felser}}, \bibinfo {author} {\bibfnamefont {I.}~\bibnamefont {Errea}}, \
  and\ \bibinfo {author} {\bibfnamefont {M.~G.}\ \bibnamefont {Vergniory}},\
  }\href {https://arxiv.org/abs/2505.09749} {\enquote {\bibinfo {title}
  {Emergent chirality and enantiomeric selectivity in layered nbox$_2$
  crystals},}\ } (\bibinfo {year} {2025}),\ \Eprint
  {http://arxiv.org/abs/2505.09749} {arXiv:2505.09749 [cond-mat.mtrl-sci]}
  \BibitemShut {NoStop}%
\bibitem [{\citenamefont {Iwasaki}\ and\ \citenamefont
  {Sugii}(1971)}]{10.1063/1.1653848}%
  \BibitemOpen
  \bibfield  {author} {\bibinfo {author} {\bibfnamefont {H.}~\bibnamefont
  {Iwasaki}}\ and\ \bibinfo {author} {\bibfnamefont {K.}~\bibnamefont
  {Sugii}},\ }\href {\doibase 10.1063/1.1653848} {\bibfield  {journal}
  {\bibinfo  {journal} {Applied Physics Letters}\ }\textbf {\bibinfo {volume}
  {19}},\ \bibinfo {pages} {92} (\bibinfo {year} {1971})}\BibitemShut {NoStop}%
\bibitem [{\citenamefont {Fava}\ \emph {et~al.}(2024)\citenamefont {Fava},
  \citenamefont {Lafargue-Dit-Hauret}, \citenamefont {Romero},\ and\
  \citenamefont {Bousquet}}]{PhysRevB.109.024113}%
  \BibitemOpen
  \bibfield  {author} {\bibinfo {author} {\bibfnamefont {M.}~\bibnamefont
  {Fava}}, \bibinfo {author} {\bibfnamefont {W.}~\bibnamefont
  {Lafargue-Dit-Hauret}}, \bibinfo {author} {\bibfnamefont {A.~H.}\
  \bibnamefont {Romero}}, \ and\ \bibinfo {author} {\bibfnamefont
  {E.}~\bibnamefont {Bousquet}},\ }\href {\doibase 10.1103/PhysRevB.109.024113}
  {\bibfield  {journal} {\bibinfo  {journal} {Phys. Rev. B}\ }\textbf {\bibinfo
  {volume} {109}},\ \bibinfo {pages} {024113} (\bibinfo {year}
  {2024})}\BibitemShut {NoStop}%
\bibitem [{\citenamefont {Wieder}(1955)}]{PhysRev.99.1161}%
  \BibitemOpen
  \bibfield  {author} {\bibinfo {author} {\bibfnamefont {H.~H.}\ \bibnamefont
  {Wieder}},\ }\href {\doibase 10.1103/PhysRev.99.1161} {\bibfield  {journal}
  {\bibinfo  {journal} {Phys. Rev.}\ }\textbf {\bibinfo {volume} {99}},\
  \bibinfo {pages} {1161} (\bibinfo {year} {1955})}\BibitemShut {NoStop}%
\bibitem [{\citenamefont {Kresse}\ and\ \citenamefont
  {Furthm\"uller}(1996)}]{vasp1}%
  \BibitemOpen
  \bibfield  {author} {\bibinfo {author} {\bibfnamefont {G.}~\bibnamefont
  {Kresse}}\ and\ \bibinfo {author} {\bibfnamefont {J.}~\bibnamefont
  {Furthm\"uller}},\ }\href {\doibase 10.1103/PhysRevB.54.11169} {\bibfield
  {journal} {\bibinfo  {journal} {Phys. Rev. B}\ }\textbf {\bibinfo {volume}
  {54}},\ \bibinfo {pages} {11169} (\bibinfo {year} {1996})}\BibitemShut
  {NoStop}%
\bibitem [{\citenamefont {Kresse}\ and\ \citenamefont
  {Furthm\"{u}ller}(1996)}]{vasp2}%
  \BibitemOpen
  \bibfield  {author} {\bibinfo {author} {\bibfnamefont {G.}~\bibnamefont
  {Kresse}}\ and\ \bibinfo {author} {\bibfnamefont {J.}~\bibnamefont
  {Furthm\"{u}ller}},\ }\href {\doibase 10.1016/0927-0256(96)00008-0}
  {\bibfield  {journal} {\bibinfo  {journal} {Comput. Mater. Sci.}\ }\textbf
  {\bibinfo {volume} {6}},\ \bibinfo {pages} {15} (\bibinfo {year}
  {1996})}\BibitemShut {NoStop}%
\bibitem [{\citenamefont {Bl\"ochl}(1994)}]{paw1}%
  \BibitemOpen
  \bibfield  {author} {\bibinfo {author} {\bibfnamefont {P.~E.}\ \bibnamefont
  {Bl\"ochl}},\ }\href {\doibase 10.1103/PhysRevB.50.17953} {\bibfield
  {journal} {\bibinfo  {journal} {Phys. Rev. B}\ }\textbf {\bibinfo {volume}
  {50}},\ \bibinfo {pages} {17953} (\bibinfo {year} {1994})}\BibitemShut
  {NoStop}%
\bibitem [{\citenamefont {Kresse}\ and\ \citenamefont {Joubert}(1999)}]{paw2}%
  \BibitemOpen
  \bibfield  {author} {\bibinfo {author} {\bibfnamefont {G.}~\bibnamefont
  {Kresse}}\ and\ \bibinfo {author} {\bibfnamefont {D.}~\bibnamefont
  {Joubert}},\ }\href {\doibase 10.1103/PhysRevB.59.1758} {\bibfield  {journal}
  {\bibinfo  {journal} {Phys. Rev. B}\ }\textbf {\bibinfo {volume} {59}},\
  \bibinfo {pages} {1758} (\bibinfo {year} {1999})}\BibitemShut {NoStop}%
\bibitem [{\citenamefont {Perdew}\ \emph {et~al.}(2008)\citenamefont {Perdew},
  \citenamefont {Ruzsinszky}, \citenamefont {Csonka}, \citenamefont {Vydrov},
  \citenamefont {Scuseria}, \citenamefont {Constantin}, \citenamefont {Zhou},\
  and\ \citenamefont {Burke}}]{PhysRevLett.100.136406}%
  \BibitemOpen
  \bibfield  {author} {\bibinfo {author} {\bibfnamefont {J.~P.}\ \bibnamefont
  {Perdew}}, \bibinfo {author} {\bibfnamefont {A.}~\bibnamefont {Ruzsinszky}},
  \bibinfo {author} {\bibfnamefont {G.~I.}\ \bibnamefont {Csonka}}, \bibinfo
  {author} {\bibfnamefont {O.~A.}\ \bibnamefont {Vydrov}}, \bibinfo {author}
  {\bibfnamefont {G.~E.}\ \bibnamefont {Scuseria}}, \bibinfo {author}
  {\bibfnamefont {L.~A.}\ \bibnamefont {Constantin}}, \bibinfo {author}
  {\bibfnamefont {X.}~\bibnamefont {Zhou}}, \ and\ \bibinfo {author}
  {\bibfnamefont {K.}~\bibnamefont {Burke}},\ }\href {\doibase
  10.1103/PhysRevLett.100.136406} {\bibfield  {journal} {\bibinfo  {journal}
  {Phys. Rev. Lett.}\ }\textbf {\bibinfo {volume} {100}},\ \bibinfo {pages}
  {136406} (\bibinfo {year} {2008})}\BibitemShut {NoStop}%
\bibitem [{si()}]{si}%
  \BibitemOpen
  \href@noop {} {\ }\bibinfo {note}
  {\href{https://journals.aps.org/prl/supplemental/10.1103/PhysRevLett.XXX.XXXXXX}{See
  Supplemental Material at }}\BibitemShut {NoStop}%
\bibitem [{\citenamefont {Perez-Mato}, \citenamefont {Orobengoa},\ and\
  \citenamefont {Aroyo}(2010)}]{Perez-Mato:sh5107}%
  \BibitemOpen
  \bibfield  {author} {\bibinfo {author} {\bibfnamefont {J.~M.}\ \bibnamefont
  {Perez-Mato}}, \bibinfo {author} {\bibfnamefont {D.}~\bibnamefont
  {Orobengoa}}, \ and\ \bibinfo {author} {\bibfnamefont {M.~I.}\ \bibnamefont
  {Aroyo}},\ }\href {\doibase 10.1107/S0108767310016247} {\bibfield  {journal}
  {\bibinfo  {journal} {Acta Crystallographica Section A}\ }\textbf {\bibinfo
  {volume} {66}},\ \bibinfo {pages} {558} (\bibinfo {year} {2010})}\BibitemShut
  {NoStop}%
\bibitem [{\citenamefont {Stokes}, \citenamefont {Hatch},\ and\ \citenamefont
  {Campbell}()}]{isotropy}%
  \BibitemOpen
  \bibfield  {author} {\bibinfo {author} {\bibfnamefont {H.~T.}\ \bibnamefont
  {Stokes}}, \bibinfo {author} {\bibfnamefont {D.~M.}\ \bibnamefont {Hatch}}, \
  and\ \bibinfo {author} {\bibfnamefont {B.~J.}\ \bibnamefont {Campbell}},\
  }\href {https://iso.byu.edu/isotropy.php} {\bibinfo  {journal} {ISOTROPY
  Software Suite}\ }\BibitemShut {NoStop}%
\bibitem [{\citenamefont {Henkelman}, \citenamefont {Uberuaga},\ and\
  \citenamefont {Jónsson}(2000)}]{10.1063/1.1329672}%
  \BibitemOpen
\bibfield  {journal} {  }\bibfield  {author} {\bibinfo {author} {\bibfnamefont
  {G.}~\bibnamefont {Henkelman}}, \bibinfo {author} {\bibfnamefont {B.~P.}\
  \bibnamefont {Uberuaga}}, \ and\ \bibinfo {author} {\bibfnamefont
  {H.}~\bibnamefont {Jónsson}},\ }\href {\doibase 10.1063/1.1329672}
  {\bibfield  {journal} {\bibinfo  {journal} {The Journal of Chemical Physics}\
  }\textbf {\bibinfo {volume} {113}},\ \bibinfo {pages} {9901} (\bibinfo {year}
  {2000})}\BibitemShut {NoStop}%
\bibitem [{\citenamefont {Heyd}, \citenamefont {Scuseria},\ and\ \citenamefont
  {Ernzerhof}(2003)}]{10.1063/1.1564060}%
  \BibitemOpen
  \bibfield  {author} {\bibinfo {author} {\bibfnamefont {J.}~\bibnamefont
  {Heyd}}, \bibinfo {author} {\bibfnamefont {G.~E.}\ \bibnamefont {Scuseria}},
  \ and\ \bibinfo {author} {\bibfnamefont {M.}~\bibnamefont {Ernzerhof}},\
  }\href {\doibase 10.1063/1.1564060} {\bibfield  {journal} {\bibinfo
  {journal} {The Journal of Chemical Physics}\ }\textbf {\bibinfo {volume}
  {118}},\ \bibinfo {pages} {8207} (\bibinfo {year} {2003})},\ \bibinfo {note}
  {erratum: $\emph{J. Chem. Phys.}$ \textbf{124}, 219906 (2006)}\BibitemShut
  {NoStop}%
\bibitem [{\citenamefont {Landau}\ and\ \citenamefont
  {Lifshitz}(1984)}]{landau2013course}%
  \BibitemOpen
  \bibfield  {author} {\bibinfo {author} {\bibfnamefont {L.~D.}\ \bibnamefont
  {Landau}}\ and\ \bibinfo {author} {\bibfnamefont {E.~M.}\ \bibnamefont
  {Lifshitz}},\ }\href@noop {} {\emph {\bibinfo {title} {Electrodynamics of
  Continuous Media, Course of Theoretical Physics}}},\ Vol.~\bibinfo {volume}
  {8}\ (\bibinfo  {publisher} {Pergamon Press, New York,},\ \bibinfo {year}
  {1984})\BibitemShut {NoStop}%
\bibitem [{\citenamefont {Zabalo}\ and\ \citenamefont
  {Stengel}(2023)}]{PhysRevLett.131.086902}%
  \BibitemOpen
  \bibfield  {author} {\bibinfo {author} {\bibfnamefont {A.}~\bibnamefont
  {Zabalo}}\ and\ \bibinfo {author} {\bibfnamefont {M.}~\bibnamefont
  {Stengel}},\ }\href {\doibase 10.1103/PhysRevLett.131.086902} {\bibfield
  {journal} {\bibinfo  {journal} {Phys. Rev. Lett.}\ }\textbf {\bibinfo
  {volume} {131}},\ \bibinfo {pages} {086902} (\bibinfo {year}
  {2023})}\BibitemShut {NoStop}%
\bibitem [{\citenamefont {Yao}, \citenamefont {Xiao},\ and\ \citenamefont
  {Niu}(2008)}]{PhysRevB.77.235406}%
  \BibitemOpen
  \bibfield  {author} {\bibinfo {author} {\bibfnamefont {W.}~\bibnamefont
  {Yao}}, \bibinfo {author} {\bibfnamefont {D.}~\bibnamefont {Xiao}}, \ and\
  \bibinfo {author} {\bibfnamefont {Q.}~\bibnamefont {Niu}},\ }\href {\doibase
  10.1103/PhysRevB.77.235406} {\bibfield  {journal} {\bibinfo  {journal} {Phys.
  Rev. B}\ }\textbf {\bibinfo {volume} {77}},\ \bibinfo {pages} {235406}
  (\bibinfo {year} {2008})}\BibitemShut {NoStop}%
\bibitem [{\citenamefont {Grieder}, \citenamefont {Tu},\ and\ \citenamefont
  {Ping}(2025)}]{https://doi.org/10.1002/adom.202501190}%
  \BibitemOpen
  \bibfield  {author} {\bibinfo {author} {\bibfnamefont {A.}~\bibnamefont
  {Grieder}}, \bibinfo {author} {\bibfnamefont {S.}~\bibnamefont {Tu}}, \ and\
  \bibinfo {author} {\bibfnamefont {Y.}~\bibnamefont {Ping}},\ }\href {\doibase
  https://doi.org/10.1002/adom.202501190} {\bibfield  {journal} {\bibinfo
  {journal} {Advanced Optical Materials}\ }\textbf {\bibinfo {volume} {13}},\
  \bibinfo {pages} {e01190} (\bibinfo {year} {2025})}\BibitemShut {NoStop}%
\end{thebibliography}%
\bibliographystyle{aipnum4-1} 

\end{document}